\documentclass[aps,prl,twocolumn,superscriptaddress,10pt]{revtex4-2}

\usepackage{amsthm}
\usepackage{amsmath,bm}
\usepackage{amssymb}
\usepackage{amsfonts}
\usepackage{graphicx}
\usepackage{txfonts}
\usepackage{xcolor}
\usepackage{float}
\usepackage{braket}

\usepackage[squaren]{SIunits}

\usepackage[colorlinks=true,linkcolor=blue,citecolor=blue,urlcolor=blue]{hyperref}

\usepackage{bbold}

\newcommand{\abs}[1]{\ensuremath{\left|{#1}\right|}}

\newcommand{\avg}[1]{\langle#1\rangle}		

\newcommand{\xzpf}{x_{\text{zpf}}}

\renewcommand{\aa}{\ensuremath{\hat{a}}}
\newcommand{\aad}{\ensuremath{\hat{a}^\dagger}}

\newcommand{\dd}{\text{d}}

\newcommand{\op}[1]{\hat{#1}}				
\newcommand{\eg}{\textit{e.g.} }				
\newcommand{\ie}{\textit{i.e.} }

\begin{document}

\title{Genuine quantum non-Gaussianity and metrological sensitivity of Fock states prepared in a mechanical resonator}

\author{Q. Rumman Rahman}
\affiliation{Department of Physics, Yale University, New Haven, 06511, CT, USA}

\author{Igor Kladarić}
\affiliation{Department of Physics, ETH Z\"{u}rich, 8093 Z\"{u}rich, Switzerland}
\affiliation{Quantum Center, ETH Z\"{u}rich, 8093 Z\"{u}rich, Switzerland}

\author{Max-Emanuel Kern}
\affiliation{Department of Physics, ETH Z\"{u}rich, 8093 Z\"{u}rich, Switzerland}
\affiliation{Quantum Center, ETH Z\"{u}rich, 8093 Z\"{u}rich, Switzerland}

\author{\\Luk\'a\v s Lachman}
\affiliation{Laboratoire Kastler Brossel, Sorbonne Universit\'e, CNRS, ENS-Universit\'e PSL, Coll\`ege de France, 4 Place Jussieu, 75005 Paris, France}
\affiliation{Department of Optics, Palack\'y University, 17. listopadu 12, 771 46 Olomouc, Czech Republic}

\author{Yiwen Chu}
\affiliation{Department of Physics, ETH Z\"{u}rich, 8093 Z\"{u}rich, Switzerland}
\affiliation{Quantum Center, ETH Z\"{u}rich, 8093 Z\"{u}rich, Switzerland}

\author{Radim Filip}
\email{filip@optics.upol.cz}
\affiliation{Department of Optics, Palack\'y University, 17. listopadu 12, 771 46 Olomouc, Czech Republic}

\author{Matteo Fadel}
\email{fadelm@phys.ethz.ch}
\affiliation{Department of Physics, ETH Z\"{u}rich, 8093 Z\"{u}rich, Switzerland}
\affiliation{Quantum Center, ETH Z\"{u}rich, 8093 Z\"{u}rich, Switzerland}

\begin{abstract}
Fock states of the quantum harmonic oscillator are fundamental to quantum sensing and information processing, serving as key resources for exploiting bosonic degrees of freedom. Here, we prepare high Fock states in a high-overtone bulk acoustic wave resonator (HBAR) by coupling it to a superconducting qubit and applying microwave pulses designed using quantum optimal control.
We characterize the experimentally realized states by employing a criterion for genuine quantum non-Gaussianity (QNG) designed to reveal multiphonon contributions. Although energy relaxation and decoherence limit the achievable fidelities, we demonstrate genuine QNG features compatible with Fock state $\ket{6}$, confirming that the prepared states cannot be generated through Gaussian operations on states with up to Fock state $\ket{5}$ contributions. 
We further investigate the robustness of these QNG features to losses and their utility in sensing displacement amplitudes. In particular, we introduce a hierarchy based on the quantum Fisher information and show that, despite decoherence and measurement imperfections, the prepared states achieve a displacement sensitivity surpassing that of an ideal Fock state $\ket{3}$. Our results have immediate applications in quantum sensing and simulations with HBAR devices.
\end{abstract}

\maketitle

Hybrid quantum devices employing mechanical degrees of freedom offer significant potential for advancing quantum technologies. 
A prominent example is circuit quantum acoustodynamics (cQAD), where solid-state mechanical resonators coupled to superconducting circuits enable the preparation, control, and readout of quantum states of mechanical motion.
This platform benefits from the availability of long-lived mechanical modes and has applications in quantum computing \cite{chamberland2022building,Qiao23}, bosonic quantum simulation \cite{BSpaper,marti2024quantum}, and quantum information storage \cite{Hann2019}. Furthermore, the relatively large effective mass of these modes makes them well-suited for precision sensing and tests of fundamental physics \cite{macroPRL,catSCI23,donadifadel24}. 

Recent experiments in cQAD have demonstrated the preparation of Schr\"odinger cat \cite{catSCI23}, squeezed \cite{marti2024quantum}, and Fock states \cite{Chu2018,vonLupke22}. 
Among these, Fock states, which are quantum states with a well-defined phonon number, are particularly desirable for sensing applications due to their non-Gaussian characteristics and fine phase-space structures \cite{wolf_motional_2019,Deng_Fock100_24}. 
Notably, Fock states exhibit invariance under phase-space rotations, making them the optimal choice for detecting displacements with unknown or random directions \cite{fadel2024reviewCV,Oh_NJP_2020,WojciechPRL22}. 
However, the generation of high-fidelity multiphonon Fock states has remained challenging, primarily due to decoherence during the extended state preparation process, which involves repeated qubit-phonon swap operations \cite{Chu2018}. As a result, while Fock states up to $\ket{7}$ have been attempted \cite{Chu2018}, the fidelity dropped below 75\% as early as $\ket{2}$ (see also \cite{vonLupke22}).

\begin{figure}[t]
    \centering
    \includegraphics[width=0.9\linewidth]{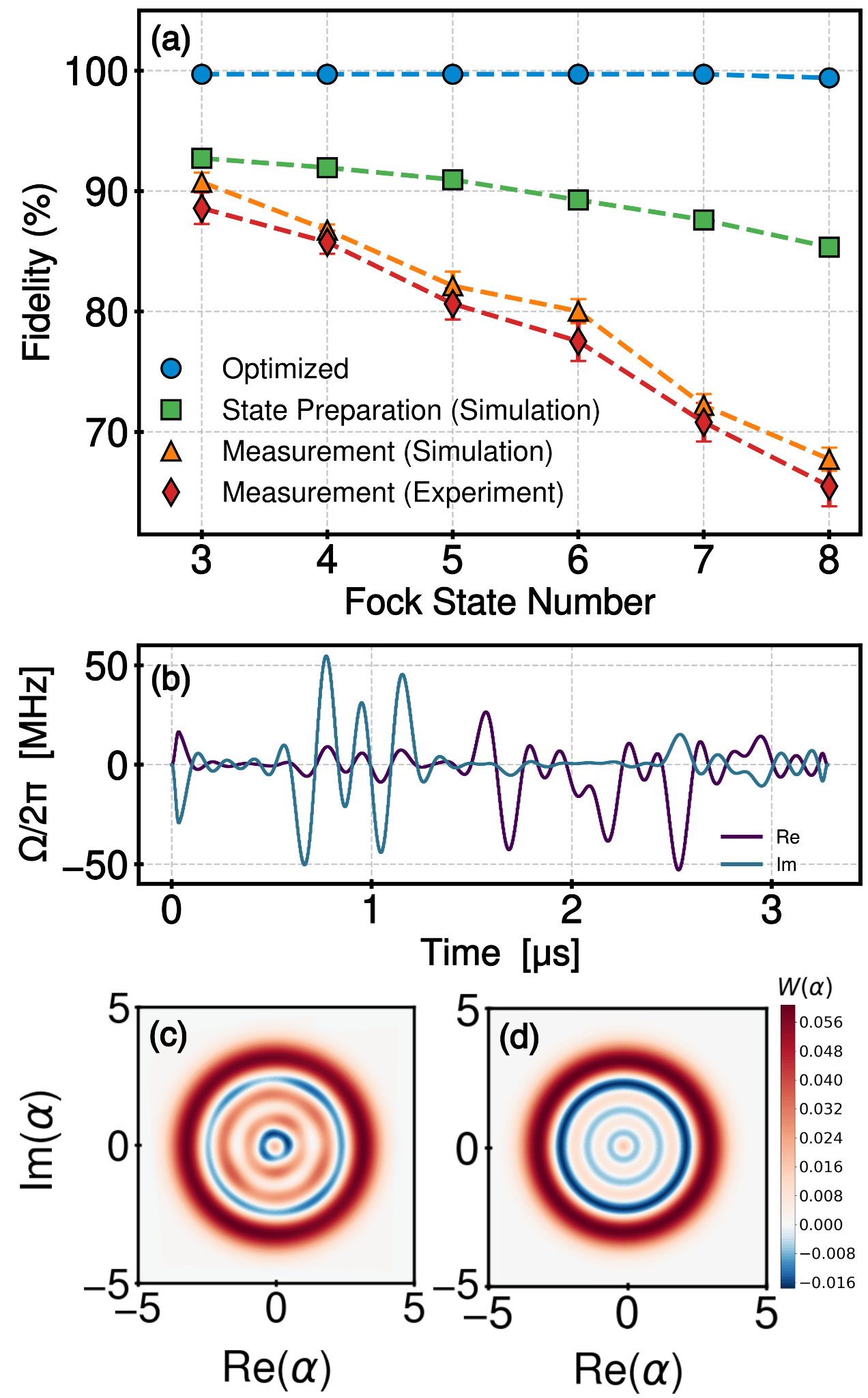}
    \caption{\textbf{Preparation of mechanical Fock states.} \textbf{(a)} fidelity of the generated pulse as achieved by the optimizer (blue), the expected fidelity of the prepared state taking into account decoherence (green), the expected measured fidelity taking into account imperfect readout (orange), and the fidelity actually measured in our experiment (red). Error bars are 1 STD, see main text for further details. 
    \textbf{(b)} real and imaginary parts of the optimal control pulse used to prepare Fock $\ket{6}$ state.
    \textbf{(c)} maximum likelihood reconstruction of the Wigner function of the state prepared using the pulse in \textbf{(b)}.
    \textbf{(d)} master-equation simulation of the Wigner function resulting from applying this pulse in the presence of decoherence and with imperfect readout.}
    \label{fig:1}
\end{figure}

Here, we apply quantum-optimal-control theory to prepare motional Fock states within a high-overtone bulk acoustic wave resonator (HBAR).  The prepared states are characterized by their fidelity, which exceeds 75\% for states up to Fock $\ket{6}$, see Fig.~\ref{fig:1}(a). Moreover, by applying criteria for genuine quantum non-Gaussianity (QNG) \cite{filip_detecting_2011,lachman_faithful_2019,podhora_quantum_2022}, we demonstrate the ability to prepare states with genuine Fock $\ket{6}$ contributions, indicating they cannot be generated by Gaussian transformations on states with up to Fock $\ket{5}$ components.
Additionally, we investigate the robustness of these non-Gaussian features under energy relaxation, which is the primary decoherence channel in our system. 
Beyond QNG, we propose a second characterization method based on the state's sensitivity to displacement amplitudes, considering experimental imperfections.
While this criterion is more stringent, it provides a direct interpretation in terms of metrological performance, demonstrating that our prepared states exhibit displacement sensitivity exceeding that of an ideal Fock $\ket{3}$ state.
For our device, this implies a quantum enhancement in force sensitivity beyond the classical limit of $\unit{63.2}{fN/\sqrt{\text{Hz}}}$.

\textbf{Experimental platform.--} The device used in this work is a cQAD system, consisting of a transmon qubit flip-chip bonded to a HBAR \cite{marti2024quantum}. The qubit operates at a frequency of $\omega_q = 2\pi\cdot\unit{5.042}{GHz}$, which can be tuned via a far-off-resonant Stark shift drive. At this frequency, the qubit has an energy relaxation time $T_1=\unit{17.2(0.4)}{\mu s}$, a Ramsey decoherence time $T_2^\ast=\unit{24.5(0.7)}{\mu s}$, and an anharmonicity $\alpha=2\pi\cdot\unit{185}{MHz}$. The HBAR is coupled to the qubit via a piezoelectric transducer made of aluminum nitride, mediating a Jaynes-Cummings (JC) interaction with a coupling strength $g=2\pi\cdot\unit{292}{kHz}$. The phonon mode we consider has a frequency $\omega_a=2\pi\cdot\unit{5.023}{GHz}$, an effective mass of about $\unit{16}{\mu g}$ \cite{catSCI23}, an energy relaxation time $T_1=\unit{89(4)}{\mu s}$ and a Ramsey decoherence time $T_2^\ast=\unit{152(9)}{\mu s}$. Our system can be described by the Hamiltonian 
\begin{align}
    \hat{H}_{\text{cQAD}}/\hbar &= \omega_q \hat{q}^\dagger \hat{q} - \dfrac{\alpha}{2} \hat{q}^{\dagger\,2} \hat{q}^2 \notag\\
    &\quad + \omega_a \aad \aa + g(\hat{q} \aad + \hat{q}^\dagger \aa) + \hat{H}_{\text{qd}}/\hbar \;, \label{eq:fullH}
\end{align}
where $\hat{q}$ and $\hat{a}$ are the bosonic annihilation operators for the qubit and the phonon mode, respectively. The term $\hat{H}_{\text{qd}}/\hbar=\Omega(t) e^{-i\omega_d t} \hat{q}^\dagger + \text{h.c.}$ describes a microwave drive at frequency $\omega_{d}$ and time-dependent amplitude $\Omega(t)$ applied to the qubit. This drive will be used for state preparation, as detailed in the following.

\vspace{2mm}
\textbf{Preparation and measurement of motional Fock states.--} 
In the strong-coupling regime, control and readout of the HBAR acoustic mode is achieved through resonant interaction with the $\{\ket{g}, \ket{e}\}$ transition of the qubit. State preparation involves tuning the qubit on resonance with the acoustic mode, \ie $\omega_q = \omega_a$, while simultaneously applying a resonant microwave drive at frequency $\omega_d = \omega_a$. The complex amplitude of the drive, $\Omega(t)$, is determined using quantum-optimal-control theory (see Fig.~\ref{fig:1}(b) for an example pulse). Our goal is to implement a unitary operation $\hat{U}_n(\Omega(t))$, mapping the initial vacuum state $\ket{0}\ket{g}$ to the desired Fock state $\ket{n}\ket{g} = \hat{U}_n(\Omega(t))\ket{0}\ket{g}$. To do this, we discretize the $I$ and $Q$ quadratures of the drive with $\unit{4}{ns}$-resolution steps and obtain them through the gradient ascent pulse engineering (GRAPE) algorithm \citep{khaneja2005optimal,eickbusch2022fast}. The optimization is performed with a target fidelity $F = |\langle n | \hat{U}(\Omega(t)) | 0 \rangle| = 0.999$. Since the closed-system solver we employ does not account for relaxation during the drive, it is essential to minimize
the effects of decoherence by using the shortest possible pulses. We do this by iteratively reducing the pulse duration until the resulting amplitude is no longer experimentally viable. Typical pulse durations range between \unit{2-4}{\mu s}, depending on the energy of the target state, in agreement with quantum speed limit estimates \cite{CanevaPRL09}.

Following state preparation, we reset the qubit by swapping any residual excited-state population with an ancillary phonon mode. The state in the original phonon mode is then measured using resonant-interaction phonon number (RPN) measurements \cite{Hofheinz2008, Chu2018}. This is achieved by initializing the qubit in the excited state, tuning it on resonance with the phonon mode of interest for a variable interaction time, and subsequently measuring the qubit population. By simulating oscillations of the qubit population during resonant interactions with phonon modes prepared in different Fock states and using them as basis functions to fit the measured qubit population, we extract the phonon number distribution of the prepared state \cite{SM}. Alternatively, displaced-parity measurements \cite{vonLupke22} can be performed using the qubit to directly measure the Wigner function. An example is presented in Fig.~\ref{fig:1}(c), which shows good agreement with the simulation prediction in Fig.~\ref{fig:1}(d). However, reconstructing Fock distributions from these measurements is delicate \cite{catSCI23,marti2024quantum}, resulting in significant uncertainties. Therefore, we rely on RPN measurements for our analyses.

Figure~\ref{fig:1}(a) illustrates the fidelities of several Fock states prepared using optimal control pulses and measured with RPN. The blue points correspond to the fidelities obtained from the Schrödinger equation-based optimizer, which assumes a closed system and perfect state readout. To model decoherence during state preparation, we employ a master equation approach that includes the effects of decay and dephasing of both the qubit and the phonon mode. These results are shown in green. To estimate the additional infidelity introduced by the readout process, we extend the master equation simulation to include the resonant interaction period between the qubit and the phonon mode during the RPN measurement. The results, depicted in orange, highlight the significant impact of energy relaxation during the measurement on the observed fidelities. Finally, we show in red the experimentally measured fidelities. The latter are in good agreement with the orange predictions, with a small systematic offset likely attributed to uncertainties in the measured decoherence rates. In addition, note that imperfections in the pulse implementation (such as amplitude miscalibrations) can result in residual qubit-phonon entanglement that will decrease the fidelity of the prepared state following qubit reset. Error bars in Fig.~\ref{fig:1}(a) represent 1 STD, computed from Monte-Carlo simulations of the fidelities (green, orange) or from the measured data (red).

We now proceed to characterize the prepared states using two distinct methods. The first employs quantum non-Gaussianity (QNG) criteria specifically designed to identify multiphonon components, while the second utilizes a metrological criterion to evaluate the displacement sensitivity of the prepared states relative to that of ideal Fock states.

\begin{figure*}[t]
    \centering
    \includegraphics[width=\textwidth]{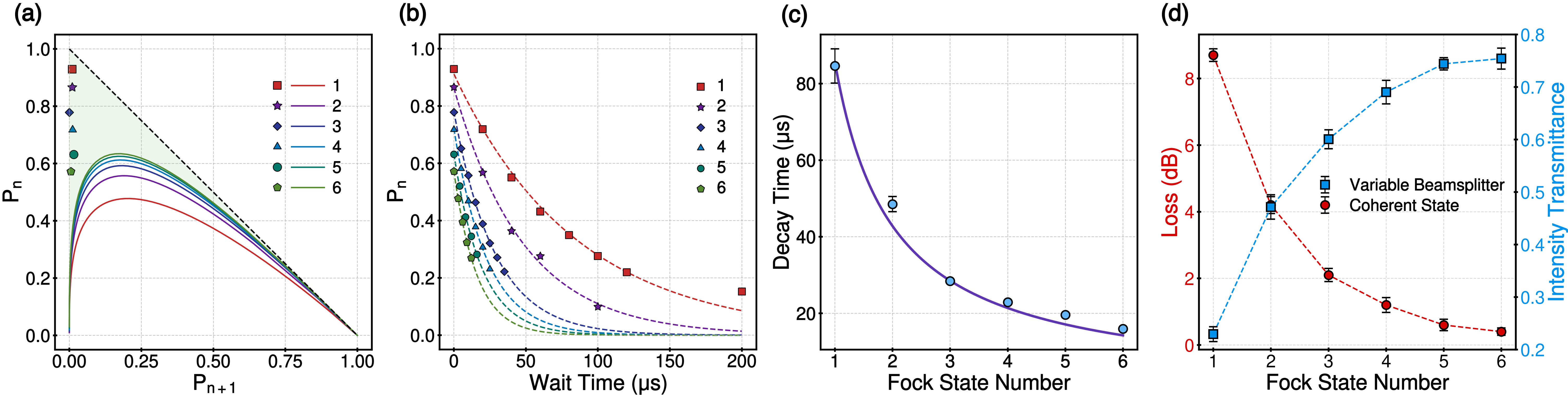}
    \caption{
    \textbf{Quantum non-Gaussianity of mechanical Fock states.}
    \textbf{(a)} experimentally measured $\{P_n,P_{n+1}\}$ for the different prepared states (colored symbols). These are compared with the associated QNG bound (lines), which is violated for our states with $n=1,..,6$. This indicates that none of the prepared states can be understood as originating from a Gaussian transformation on a state composed by Fock states up to $n-1$. 
    Note that only the region below the dashed diagonal line has $P_n+P_{n+1}\leq 1$ and is thus physical.
    \textbf{(b)} measurement of the decay of $P_n$ as a function of the waiting time between state preparation and measurement. Dashed lines are exponential fits to the data.
    Decay times resulting from the fits are reported in \textbf{(c)}, together with a $\tau_1/n$ function (purple line) where $\tau_1=\unit{85(4)}{\mu s}$ is the decay time for Fock $\ket{1}$. We observe the expected behaviour that a Fock $\ket{n}$ state decays $n$ times faster than Fock $\ket{1}$. \textbf{(d)} robustness of the prepared states in terms of the amount of losses tolerated before the corresponding QNG bound stops being violated (see main text).
    }
    \label{fig:2}
\end{figure*}

\vspace{2mm}
\textbf{QNG characterization.--} While the negativity of the Wigner function is indicative of non-classical features, it alone does not provide a quantitative measure of quantum non-Gaussianity (QNG). To address this, QNG criteria have been developed to provide hierarchical conditions for reliably identifying states generated by genuinely nonlinear dynamics and for quantifying their degree of non-Gaussianity \citep{lachman_faithful_2019, podhora_quantum_2022, filip_detecting_2011}.

A pure state shows genuine $n$-phonon QNG if it cannot be expressed as a superposition of Fock states up to $\ket{n-1}$, nor as any Gaussian transformations applied to such states.
Namely, a pure state $\ket{\psi_n}$ shows genuine $n$-phonon QNG if
\begin{equation}\label{eq:QNGcond}
    \ket{\psi_n} \neq \ket{\psi_{n-1}(\alpha,r,\{c_m\})} \equiv \hat{D}(\alpha) \hat{S}(r) \sum_{m=0}^{n-1} c_m \ket{m}
\end{equation}
for all $\alpha$, $r$, and $\{c_m\}$ such that $\sum_m|c_m|^2=1$. Here, $\hat{D}(\alpha)$ and $\hat{S}(r)$ represent the displacement and squeezing operators, respectively, parametrized by the complex displacement amplitude $\alpha$ and squeezing parameter $r$. 
To ensure experimental applicability, Eq.~\eqref{eq:QNGcond} needs to be generalized to mixed states \(\rho\).
A mixed state $\rho_n$ shows genuine $n$-phonon QNG if
\begin{equation}\label{eq:QNGmixed}
\rho_n \neq \int P(r, \alpha, \{c_m\}) \ket{\psi_{n-1}(\alpha,r,\{c_m\})}\bra{\psi_{n-1}(\alpha,r,\{c_m\})} \, \dd^2\alpha \, \dd^2r \;,
\end{equation}
where $P(r, \alpha, \{c_m\})$ is the probability distribution characterizing the statistical mixture of pure states.

From Eq.~\eqref{eq:QNGcond}, a QNG criterion can be derived by analyzing the probability \(P_n\) of the state being measured in the Fock state \(\ket{n}\). Genuine $n$-phonon QNG is revealed if \(P_n\) exceeds the threshold $\overline{P}_n \equiv \max_{\alpha, r, \{c_m\}} \left| \langle n \ket{\psi_{n-1}(\alpha,r,\{c_m\})} \right|^2$ \cite{podhora_quantum_2022,chabaud_certification_2021,Fiurasek_22}.
Here, the linearity of the maximization ensures that no convex combination of excluded states can satisfy this requirement, allowing the criterion to be extended to mixed states. 
However, the threshold $\overline{P}_n$ is often too stringent for mixed states, as its violation generally demands high state purity.

In practical settings, quantum states are affected by decoherence and noise, which complicate the detection of QNG. 
Specifically, criteria based solely on the measurement of \( P_n \) are highly sensitive to energy relaxation and detection inefficiency, and may quickly fail to reveal non-Gaussian features. When losses prevent surpassing the absolute threshold $\overline{P}_n$, a more robust hierarchical criterion is needed. To this end, we extend the approach in Ref.~\cite{Provaznik20}, which introduces a functional \( F_{a, n}(\rho) \equiv P_n + aP_{n+1} \) to robustly identify genuine \( n \)-phonon QNG in the presence of experimental imperfections. Here, \( P_n \) denotes the probability of detecting \( n \) phonons (the ``success" probability), \( P_{n+1} \) represents the probability of detecting at least \( n+1 \) phonons (the ``error" probability), and \( a \) is a free parameter. The threshold for this criterion, $\overline{F}_n(a)$, is obtained by maximizing \( F_{a, n}(\rho) \) over all states described by the right-hand side of Eq.~\eqref{eq:QNGmixed}. If the experimentally observed \( F_{a, n}(\rho) \) exceeds the threshold $\overline{F}_n(a)$ for some value of \( a \), we can conclude that the state exhibits genuine \( n \)-phonon QNG. 

For a given $n$, $\overline{F}_n(a)$ can be represented in the $\{P_n,P_{n+1}\}$-plane as a curve parametrized by $a$, see Fig.~\ref{fig:2}(a).
Note that the maximum of this curve over $P_{n+1}$ coincides with the threshold $\overline{P}_n$, showing that the criterion relying solely on the measurement of $P_n$ is significantly more stringent. The advantage of using information about $P_{n+1}$ can be intuitively understood from the fact that QNG manifests as sharp variations in the Fock state distribution. Since sharp differences between $P_n$ and $P_{n-1}$ are diminished by losses, a criterion based on $P_n$ and $P_{n+1}$ is more robust to this imperfection. Moreover, by noting that $\overline{F}_n(a)$ can in principle be violated for small $P_n$, it is now evident that the detection of QNG does not necessarily require high fidelity or purity of the prepared states.

To characterize the experimentally prepared Fock states, we plot the measured values of \(\{P_n, P_{n+1}\}\) in Fig.~\ref{fig:2}(a), demonstrating that they violate the QNG thresholds \(\overline{F}_n(a)\) for \(n \leq 6\). 
Measurements for $n\geq 7$ did not show a violation of the corresponding threshold, likely due to an increased \(P_{n+1}\) population caused by imperfections in the state preparation pulse. In the following, we further investigate the robustness of the QNG features against losses.

\textbf{QNG depth.--}
For each of our experimentally demonstrated non-Gaussian states, we aim to quantify the maximum amount of losses they can tolerate before ceasing to violate the corresponding QNG threshold, see Fig.~\ref{fig:2}(a). 
This provides a measure of the robustness of the observed non-Gaussian features, also referred to as the QNG depth \cite{Lee_depth_PRA91,Lee_depth_PRA92,Radim_depth_14}. To achieve this, we first prepare the desired state as before; however, we now allow the phonon mode to evolve for a variable wait time before performing the RPN measurement. During this wait time, the qubit is detuned far from the mode to suppress additional losses. The resulting $P_n$ population decay over time, shown in Fig.~\ref{fig:2}(b), follows the expected exponential decay. By fitting each dataset with the function $e^{-t/\tau_n}$, we extract the decay times $\tau_n$, which agree with the expected dependence $\tau_n = \tau_1/n$, see Fig.~\ref{fig:2}(c). Notably, the $\tau_1$ value measured here using RPN aligns with the independently measured $T_1$.

Next, for each prepared state, we identify the wait time $\tau_n^\ast$ at which the observed $\{P_n, P_{n+1}\}$ values intersect the corresponding QNG threshold. This time, $\tau_n^\ast$ represents the maximum loss the state can tolerate before it no longer exhibits genuine $n$-phonon QNG. To quantify this tolerance independently of the specifics of the current experiment and enable comparisons with other platforms, we use two independent methods. First, we perform a master equation simulation with collapse operator $\sqrt{1/\tau_1}a$, where a coherent state of fixed amplitude is evolved for a time equal to the maximum wait time identified for each state $\tau_n^\ast$. The tolerable loss for each state, expressed in dB, is then calculated as $-10 \log_{10}(\abs{\alpha_\text{fin}}^2/\abs{\alpha_\text{in}}^2)$, where $\alpha_{\text{in}(\text{out})}$ is the initial (final) amplitude of this coherent state. Second, we simulate the action of a beamsplitter with an adjustable splitting ratio on the initial state measured at zero wait time. The beamsplitter's transmittance is adjusted to model the equivalent loss corresponding to the simulated free-evolution duration. By determining the minimum transmittance required to still violate the QNG threshold, we quantify the robustness of the state under loss. The results of both analyses are presented in Fig.~\ref{fig:2}(d). 
As expected, higher Fock states are more sensitive to losses.

\vspace{2mm}
\textbf{Fisher information for displacement sensing.--} 
We now investigate how effectively the states we have prepared perform in the metrological task of estimating the amplitude $d=\abs{\alpha}$ of a displacement.
The upper bound for sensitivity is given by the quantum Cramér-Rao bound $(\Delta d)^2 \geq 1/\nu F_Q[\op{\rho}, \op{G}]$, where $\nu$ is the number of measurements and $F_Q[\op{\rho}, \op{G}]$ is the quantum Fisher information (QFI) of the state $\op{\rho}$ with respect to the generator $\op{G}$. 
For a pure state undergoing unitary evolution, the QFI is $F_Q[\op{\rho}, \op{G}]=4\text{Var}[\op{G}]_{\op{\rho}}$. In the case of displacements, $\op{G}=\aa e^{i\phi}- \aad e^{-i\phi}$ corresponds to the phase space quadrature perpendicular to the displacement direction. 
If the displacement direction is unknown, Fock states are optimal for sensing and achieve $F_Q[\ket{n}, \op{G}]=4(2n+1)$ \cite{SM,fadel2024reviewCV}. By measuring the populations of the state's Fock basis, one obtains a probability distribution $P_m(d)=\bra{m}e^{-i d \hat{G}}\op{\rho} e^{i d \hat{G}}\ket{m}$ for observing outcome $m$. The sensitivity of estimating $d$ from this distribution depends on the classical Fisher information (FI) given by $F[\op{\rho},\op{G},\{\ket{m}\}]=\sum_m \frac{1}{P_m(d)}\left(\frac{\partial P_m(d)}{\partial d}\right)^2$. 
Importantly, for displacement sensing with Fock states, the classical Fisher information satisfies $F[\ket{n},\op{G},\{\ket{m}\}]=F_Q[\ket{n},\op{G}]$, meaning that population measurements are optimal.

\begin{figure}[t]
    \centering
    \includegraphics[width=0.9\linewidth]{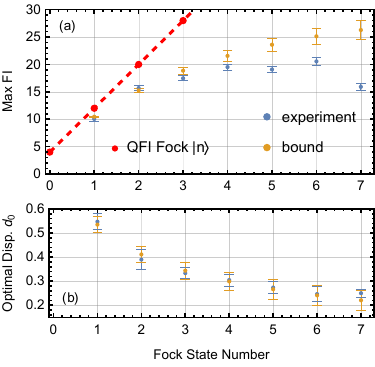}
    \caption{\textbf{Displacement sensing.} \textbf{(a)} QFI $4(2n+1)$ for displacement sensing with a Fock state $\ket{n}$ (red), compared to the FI computed for the states we have prepared (blue) and to the simulation of the FI attained by a Fock state after having taken into account decoherence in our system (yellow). 
    \textbf{(b)} displacement amplitudes $d_0$ that maximize the FI for the experimental (blue) and simulated (yellow) states.}
    \label{fig:3}
\end{figure}

Since for the ground state $\ket{0}$, as well as for any other coherent state, $F_Q[\ket{n},\op{G}]=4$, observing a FI larger than 4 indicates a nonclassical metrological advantage. Following this idea, a comparison with the Fock state bounds $F_Q[\ket{n},\op{G}]=4(2n+1)$ provides a quantification of the state's metrological performance. In Figure~\ref{fig:3}(a) we illustrate these theoretical bounds (red), alongside the maximum FI of our prepared states (blue). 
The displacement amplitude $d_0$ achieving this maximum value is shown in Fig.~\ref{fig:3}(b). We observe that the experimentally achieved sensitivity for all prepared states does not exceed the theoretical bound for the Fock state $\ket{2}$.

Note that, in a realistic experiment, even if the state preparation achieves unit fidelity, any imperfection in the measurement may result in the impossibility to observe the expected sensitivity. For this reason, we propose to take into account the effect of measurement imperfections and adjust the metrological hierarchy accordingly. In our case, the main limitation comes from the long RPN measurement time of $\unit{\sim 10}{\mu s}$, during which energy relaxation of the state occurs. We thus model our measurement step as a decay process with time constant $T_1$, from which we simulate what would be the effect of measuring a Fock state $\ket{n}$. This can be understood as an amplitude-damping quantum channel $\Lambda[\hat{\rho}]$, whose action on the state $\hat{\rho}$ can be computed analytically \cite{SM}. The result is a hierarchy $F[\Lambda[\ket{n}],\op{G},\{\ket{m}\}]$, which is illustrated by the yellow points in Fig.~\ref{fig:3}(a). Taking this characterization of the device into account, we can conclude that the prepared states with $n=6$ can offer a sensitivity better than that of the Fock state $\ket{3}$.
This implies that our device can achieve a quantum-enhanced force sensitivity beyond the classical limit of $\unit{63.2}{fN/\sqrt{\text{Hz}}}$ \cite{SM}.

\vspace{2mm}
\textbf{Conclusions.--} We have investigated the preparation and characterization of Fock states with high phonon number in a mechanical oscillator. 
Using optimal control, we have designed pulses optimized for preparing Fock states in our HBAR device coupled to a superconducting qubit. 
We then implemented these pulses experimentally and measured the resulting phonon number distribution to evaluate the quality of the prepared states, focusing on their quantum non-Gaussianity and metrological performance. 
We demonstrate the ability to prepare states with genuine Fock $\ket{6}$ contributions, as well as with a displacement sensitivity higher than that of an ideal Fock $\ket{3}$ state under the same lossy dynamics.
Our results highlight the relation between QNG and metrological resources \cite{KwonPRL19}, and have immediate applications in  quantum sensing of forces and fields  \cite{fadel2024reviewCV,Deng_Fock100_24,wolf_motional_2019}, tests of fundamental physics \cite{macroPRL,donadifadel24} and simulations with multiphonon Fock states \cite{sturges_quantum_2021}.

\vspace{2mm}
\textit{Acknowledgments.--} 
Q.R.R. was supported by the Alan S. Tetelman 1958 Fellowships for International Research in the Sciences, administered by Yale University.
L.L. acknowledges the support from project No. 23-06015O of the Czech Science Foundation.
R.F. acknowledges the grant No. 21-13265X of the Czech Science Foundation.
M.F. was supported by the Swiss National Science Foundation Ambizione Grant No. 208886, and by The Branco Weiss Fellowship -- Society in Science, administered by the ETH Z\"{u}rich.

\bibliographystyle{apsrev4-1} 
\bibliography{mybib.bib}

\begin{thebibliography}{32}%
\makeatletter
\providecommand \@ifxundefined [1]{%
 \@ifx{#1\undefined}
}%
\providecommand \@ifnum [1]{%
 \ifnum #1\expandafter \@firstoftwo
 \else \expandafter \@secondoftwo
 \fi
}%
\providecommand \@ifx [1]{%
 \ifx #1\expandafter \@firstoftwo
 \else \expandafter \@secondoftwo
 \fi
}%
\providecommand \natexlab [1]{#1}%
\providecommand \enquote  [1]{``#1''}%
\providecommand \bibnamefont  [1]{#1}%
\providecommand \bibfnamefont [1]{#1}%
\providecommand \citenamefont [1]{#1}%
\providecommand \href@noop [0]{\@secondoftwo}%
\providecommand \href [0]{\begingroup \@sanitize@url \@href}%
\providecommand \@href[1]{\@@startlink{#1}\@@href}%
\providecommand \@@href[1]{\endgroup#1\@@endlink}%
\providecommand \@sanitize@url [0]{\catcode `\\12\catcode `\$12\catcode `\&12\catcode `\#12\catcode `\^12\catcode `\_12\catcode `\%12\relax}%
\providecommand \@@startlink[1]{}%
\providecommand \@@endlink[0]{}%
\providecommand \url  [0]{\begingroup\@sanitize@url \@url }%
\providecommand \@url [1]{\endgroup\@href {#1}{\urlprefix }}%
\providecommand \urlprefix  [0]{URL }%
\providecommand \Eprint [0]{\href }%
\providecommand \doibase [0]{http://dx.doi.org/}%
\providecommand \selectlanguage [0]{\@gobble}%
\providecommand \bibinfo  [0]{\@secondoftwo}%
\providecommand \bibfield  [0]{\@secondoftwo}%
\providecommand \translation [1]{[#1]}%
\providecommand \BibitemOpen [0]{}%
\providecommand \bibitemStop [0]{}%
\providecommand \bibitemNoStop [0]{.\EOS\space}%
\providecommand \EOS [0]{\spacefactor3000\relax}%
\providecommand \BibitemShut  [1]{\csname bibitem#1\endcsname}%
\let\auto@bib@innerbib\@empty
\bibitem [{\citenamefont {Chamberland}\ \emph {et~al.}(2022)\citenamefont {Chamberland}, \citenamefont {Noh}, \citenamefont {Arrangoiz-Arriola}, \citenamefont {Campbell}, \citenamefont {Hann}, \citenamefont {Iverson}, \citenamefont {Putterman}, \citenamefont {Bohdanowicz}, \citenamefont {Flammia}, \citenamefont {Keller} \emph {et~al.}}]{chamberland2022building}%
  \BibitemOpen
  \bibfield  {author} {\bibinfo {author} {\bibfnamefont {C.}~\bibnamefont {Chamberland}}, \bibinfo {author} {\bibfnamefont {K.}~\bibnamefont {Noh}}, \bibinfo {author} {\bibfnamefont {P.}~\bibnamefont {Arrangoiz-Arriola}}, \bibinfo {author} {\bibfnamefont {E.~T.}\ \bibnamefont {Campbell}}, \bibinfo {author} {\bibfnamefont {C.~T.}\ \bibnamefont {Hann}}, \bibinfo {author} {\bibfnamefont {J.}~\bibnamefont {Iverson}}, \bibinfo {author} {\bibfnamefont {H.}~\bibnamefont {Putterman}}, \bibinfo {author} {\bibfnamefont {T.~C.}\ \bibnamefont {Bohdanowicz}}, \bibinfo {author} {\bibfnamefont {S.~T.}\ \bibnamefont {Flammia}}, \bibinfo {author} {\bibfnamefont {A.}~\bibnamefont {Keller}},  \emph {et~al.},\ }\href {\doibase http://dx.doi.org/10.1103/PRXQuantum.3.010329} {\bibfield  {journal} {\bibinfo  {journal} {PRX Quantum}\ }\textbf {\bibinfo {volume} {3}},\ \bibinfo {pages} {010329} (\bibinfo {year} {2022})}\BibitemShut {NoStop}%
\bibitem [{\citenamefont {Qiao}\ \emph {et~al.}(2023)\citenamefont {Qiao}, \citenamefont {Dumur}, \citenamefont {Andersson}, \citenamefont {Yan}, \citenamefont {Chou}, \citenamefont {Grebel}, \citenamefont {Conner}, \citenamefont {Joshi}, \citenamefont {Miller}, \citenamefont {Povey}, \citenamefont {Wu},\ and\ \citenamefont {Cleland}}]{Qiao23}%
  \BibitemOpen
  \bibfield  {author} {\bibinfo {author} {\bibfnamefont {H.}~\bibnamefont {Qiao}}, \bibinfo {author} {\bibfnamefont {E.}~\bibnamefont {Dumur}}, \bibinfo {author} {\bibfnamefont {G.}~\bibnamefont {Andersson}}, \bibinfo {author} {\bibfnamefont {H.}~\bibnamefont {Yan}}, \bibinfo {author} {\bibfnamefont {M.-H.}\ \bibnamefont {Chou}}, \bibinfo {author} {\bibfnamefont {J.}~\bibnamefont {Grebel}}, \bibinfo {author} {\bibfnamefont {C.~R.}\ \bibnamefont {Conner}}, \bibinfo {author} {\bibfnamefont {Y.~J.}\ \bibnamefont {Joshi}}, \bibinfo {author} {\bibfnamefont {J.~M.}\ \bibnamefont {Miller}}, \bibinfo {author} {\bibfnamefont {R.~G.}\ \bibnamefont {Povey}}, \bibinfo {author} {\bibfnamefont {X.}~\bibnamefont {Wu}}, \ and\ \bibinfo {author} {\bibfnamefont {A.~N.}\ \bibnamefont {Cleland}},\ }\href {\doibase 10.1126/science.adg8715} {\bibfield  {journal} {\bibinfo  {journal} {Science}\ }\textbf {\bibinfo {volume} {380}},\ \bibinfo {pages} {1030} (\bibinfo {year} {2023})}\BibitemShut {NoStop}%
\bibitem [{\citenamefont {von L{\"u}pke}\ \emph {et~al.}(2024)\citenamefont {von L{\"u}pke}, \citenamefont {Rodrigues}, \citenamefont {Yang}, \citenamefont {Fadel},\ and\ \citenamefont {Chu}}]{BSpaper}%
  \BibitemOpen
  \bibfield  {author} {\bibinfo {author} {\bibfnamefont {U.}~\bibnamefont {von L{\"u}pke}}, \bibinfo {author} {\bibfnamefont {I.~C.}\ \bibnamefont {Rodrigues}}, \bibinfo {author} {\bibfnamefont {Y.}~\bibnamefont {Yang}}, \bibinfo {author} {\bibfnamefont {M.}~\bibnamefont {Fadel}}, \ and\ \bibinfo {author} {\bibfnamefont {Y.}~\bibnamefont {Chu}},\ }\href {https://doi.org/10.1038/s41567-023-02377-w} {\bibfield  {journal} {\bibinfo  {journal} {Nature Physics}\ }\textbf {\bibinfo {volume} {20}},\ \bibinfo {pages} {564} (\bibinfo {year} {2024})}\BibitemShut {NoStop}%
\bibitem [{\citenamefont {Marti}\ \emph {et~al.}(2024)\citenamefont {Marti}, \citenamefont {von L{\"u}pke}, \citenamefont {Joshi}, \citenamefont {Yang}, \citenamefont {Bild}, \citenamefont {Omahen}, \citenamefont {Chu},\ and\ \citenamefont {Fadel}}]{marti2024quantum}%
  \BibitemOpen
  \bibfield  {author} {\bibinfo {author} {\bibfnamefont {S.}~\bibnamefont {Marti}}, \bibinfo {author} {\bibfnamefont {U.}~\bibnamefont {von L{\"u}pke}}, \bibinfo {author} {\bibfnamefont {O.}~\bibnamefont {Joshi}}, \bibinfo {author} {\bibfnamefont {Y.}~\bibnamefont {Yang}}, \bibinfo {author} {\bibfnamefont {M.}~\bibnamefont {Bild}}, \bibinfo {author} {\bibfnamefont {A.}~\bibnamefont {Omahen}}, \bibinfo {author} {\bibfnamefont {Y.}~\bibnamefont {Chu}}, \ and\ \bibinfo {author} {\bibfnamefont {M.}~\bibnamefont {Fadel}},\ }\href {\doibase 10.1038/s41567-024-02545-6} {\bibfield  {journal} {\bibinfo  {journal} {Nature Physics}\ }\textbf {\bibinfo {volume} {20}},\ \bibinfo {pages} {1448} (\bibinfo {year} {2024})}\BibitemShut {NoStop}%
\bibitem [{\citenamefont {Hann}\ \emph {et~al.}(2019)\citenamefont {Hann}, \citenamefont {Zou}, \citenamefont {Zhang}, \citenamefont {Chu}, \citenamefont {Schoelkopf}, \citenamefont {Girvin},\ and\ \citenamefont {Jiang}}]{Hann2019}%
  \BibitemOpen
  \bibfield  {author} {\bibinfo {author} {\bibfnamefont {C.~T.}\ \bibnamefont {Hann}}, \bibinfo {author} {\bibfnamefont {C.-L.}\ \bibnamefont {Zou}}, \bibinfo {author} {\bibfnamefont {Y.}~\bibnamefont {Zhang}}, \bibinfo {author} {\bibfnamefont {Y.}~\bibnamefont {Chu}}, \bibinfo {author} {\bibfnamefont {R.~J.}\ \bibnamefont {Schoelkopf}}, \bibinfo {author} {\bibfnamefont {S.~M.}\ \bibnamefont {Girvin}}, \ and\ \bibinfo {author} {\bibfnamefont {L.}~\bibnamefont {Jiang}},\ }\href {\doibase 10.1103/PhysRevLett.123.250501} {\bibfield  {journal} {\bibinfo  {journal} {Phys. Rev. Lett.}\ }\textbf {\bibinfo {volume} {123}},\ \bibinfo {pages} {250501} (\bibinfo {year} {2019})}\BibitemShut {NoStop}%
\bibitem [{\citenamefont {Schrinski}\ \emph {et~al.}(2023)\citenamefont {Schrinski}, \citenamefont {Yang}, \citenamefont {von L\"upke}, \citenamefont {Bild}, \citenamefont {Chu}, \citenamefont {Hornberger}, \citenamefont {Nimmrichter},\ and\ \citenamefont {Fadel}}]{macroPRL}%
  \BibitemOpen
  \bibfield  {author} {\bibinfo {author} {\bibfnamefont {B.}~\bibnamefont {Schrinski}}, \bibinfo {author} {\bibfnamefont {Y.}~\bibnamefont {Yang}}, \bibinfo {author} {\bibfnamefont {U.}~\bibnamefont {von L\"upke}}, \bibinfo {author} {\bibfnamefont {M.}~\bibnamefont {Bild}}, \bibinfo {author} {\bibfnamefont {Y.}~\bibnamefont {Chu}}, \bibinfo {author} {\bibfnamefont {K.}~\bibnamefont {Hornberger}}, \bibinfo {author} {\bibfnamefont {S.}~\bibnamefont {Nimmrichter}}, \ and\ \bibinfo {author} {\bibfnamefont {M.}~\bibnamefont {Fadel}},\ }\href {\doibase 10.1103/PhysRevLett.130.133604} {\bibfield  {journal} {\bibinfo  {journal} {Phys. Rev. Lett.}\ }\textbf {\bibinfo {volume} {130}},\ \bibinfo {pages} {133604} (\bibinfo {year} {2023})}\BibitemShut {NoStop}%
\bibitem [{\citenamefont {Bild}\ \emph {et~al.}(2023)\citenamefont {Bild}, \citenamefont {Fadel}, \citenamefont {Yang}, \citenamefont {von L\"upke}, \citenamefont {Martin}, \citenamefont {Bruno},\ and\ \citenamefont {Chu}}]{catSCI23}%
  \BibitemOpen
  \bibfield  {author} {\bibinfo {author} {\bibfnamefont {M.}~\bibnamefont {Bild}}, \bibinfo {author} {\bibfnamefont {M.}~\bibnamefont {Fadel}}, \bibinfo {author} {\bibfnamefont {Y.}~\bibnamefont {Yang}}, \bibinfo {author} {\bibfnamefont {U.}~\bibnamefont {von L\"upke}}, \bibinfo {author} {\bibfnamefont {P.}~\bibnamefont {Martin}}, \bibinfo {author} {\bibfnamefont {A.}~\bibnamefont {Bruno}}, \ and\ \bibinfo {author} {\bibfnamefont {Y.}~\bibnamefont {Chu}},\ }\href {\doibase 10.1126/science.adf7553} {\bibfield  {journal} {\bibinfo  {journal} {Science}\ }\textbf {\bibinfo {volume} {380}},\ \bibinfo {pages} {274} (\bibinfo {year} {2023})}\BibitemShut {NoStop}%
\bibitem [{\citenamefont {Donadi}\ and\ \citenamefont {Fadel}(2025)}]{donadifadel24}%
  \BibitemOpen
  \bibfield  {author} {\bibinfo {author} {\bibfnamefont {S.}~\bibnamefont {Donadi}}\ and\ \bibinfo {author} {\bibfnamefont {M.}~\bibnamefont {Fadel}},\ }\href {\doibase 10.1103/PhysRevD.111.026009} {\bibfield  {journal} {\bibinfo  {journal} {Phys. Rev. D}\ }\textbf {\bibinfo {volume} {111}},\ \bibinfo {pages} {026009} (\bibinfo {year} {2025})}\BibitemShut {NoStop}%
\bibitem [{\citenamefont {Chu}\ \emph {et~al.}(2018)\citenamefont {Chu}, \citenamefont {Kharel}, \citenamefont {Yoon}, \citenamefont {Frunzio}, \citenamefont {Rakich},\ and\ \citenamefont {Schoelkopf}}]{Chu2018}%
  \BibitemOpen
  \bibfield  {author} {\bibinfo {author} {\bibfnamefont {Y.}~\bibnamefont {Chu}}, \bibinfo {author} {\bibfnamefont {P.}~\bibnamefont {Kharel}}, \bibinfo {author} {\bibfnamefont {T.}~\bibnamefont {Yoon}}, \bibinfo {author} {\bibfnamefont {L.}~\bibnamefont {Frunzio}}, \bibinfo {author} {\bibfnamefont {P.~T.}\ \bibnamefont {Rakich}}, \ and\ \bibinfo {author} {\bibfnamefont {R.~J.}\ \bibnamefont {Schoelkopf}},\ }\href {https://doi.org/10.1038/s41586-018-0717-7} {\bibfield  {journal} {\bibinfo  {journal} {Nature}\ }\textbf {\bibinfo {volume} {563}},\ \bibinfo {pages} {666} (\bibinfo {year} {2018})}\BibitemShut {NoStop}%
\bibitem [{\citenamefont {von L{\"u}pke}\ \emph {et~al.}(2022)\citenamefont {von L{\"u}pke}, \citenamefont {Yang}, \citenamefont {Bild}, \citenamefont {Michaud}, \citenamefont {Fadel},\ and\ \citenamefont {Chu}}]{vonLupke22}%
  \BibitemOpen
  \bibfield  {author} {\bibinfo {author} {\bibfnamefont {U.}~\bibnamefont {von L{\"u}pke}}, \bibinfo {author} {\bibfnamefont {Y.}~\bibnamefont {Yang}}, \bibinfo {author} {\bibfnamefont {M.}~\bibnamefont {Bild}}, \bibinfo {author} {\bibfnamefont {L.}~\bibnamefont {Michaud}}, \bibinfo {author} {\bibfnamefont {M.}~\bibnamefont {Fadel}}, \ and\ \bibinfo {author} {\bibfnamefont {Y.}~\bibnamefont {Chu}},\ }\href {\doibase 10.1038/s41567-022-01591-2} {\bibfield  {journal} {\bibinfo  {journal} {Nature Physics}\ }\textbf {\bibinfo {volume} {18}},\ \bibinfo {pages} {794} (\bibinfo {year} {2022})}\BibitemShut {NoStop}%
\bibitem [{\citenamefont {Wolf}\ \emph {et~al.}(2019)\citenamefont {Wolf}, \citenamefont {Shi}, \citenamefont {Heip}, \citenamefont {Gessner}, \citenamefont {Pezzè}, \citenamefont {Smerzi}, \citenamefont {Schulte}, \citenamefont {Hammerer},\ and\ \citenamefont {Schmidt}}]{wolf_motional_2019}%
  \BibitemOpen
  \bibfield  {author} {\bibinfo {author} {\bibfnamefont {F.}~\bibnamefont {Wolf}}, \bibinfo {author} {\bibfnamefont {C.}~\bibnamefont {Shi}}, \bibinfo {author} {\bibfnamefont {J.~C.}\ \bibnamefont {Heip}}, \bibinfo {author} {\bibfnamefont {M.}~\bibnamefont {Gessner}}, \bibinfo {author} {\bibfnamefont {L.}~\bibnamefont {Pezzè}}, \bibinfo {author} {\bibfnamefont {A.}~\bibnamefont {Smerzi}}, \bibinfo {author} {\bibfnamefont {M.}~\bibnamefont {Schulte}}, \bibinfo {author} {\bibfnamefont {K.}~\bibnamefont {Hammerer}}, \ and\ \bibinfo {author} {\bibfnamefont {P.~O.}\ \bibnamefont {Schmidt}},\ }\href {\doibase 10.1038/s41467-019-10576-4} {\bibfield  {journal} {\bibinfo  {journal} {Nat Commun}\ }\textbf {\bibinfo {volume} {10}},\ \bibinfo {pages} {2929} (\bibinfo {year} {2019})}\BibitemShut {NoStop}%
\bibitem [{\citenamefont {Deng}\ \emph {et~al.}(2024)\citenamefont {Deng}, \citenamefont {Li}, \citenamefont {Chen}, \citenamefont {Ni}, \citenamefont {Cai}, \citenamefont {Mai}, \citenamefont {Zhang}, \citenamefont {Zheng}, \citenamefont {Yu}, \citenamefont {Zou}, \citenamefont {Liu}, \citenamefont {Yan}, \citenamefont {Xu},\ and\ \citenamefont {Yu}}]{Deng_Fock100_24}%
  \BibitemOpen
  \bibfield  {author} {\bibinfo {author} {\bibfnamefont {X.}~\bibnamefont {Deng}}, \bibinfo {author} {\bibfnamefont {S.}~\bibnamefont {Li}}, \bibinfo {author} {\bibfnamefont {Z.-J.}\ \bibnamefont {Chen}}, \bibinfo {author} {\bibfnamefont {Z.}~\bibnamefont {Ni}}, \bibinfo {author} {\bibfnamefont {Y.}~\bibnamefont {Cai}}, \bibinfo {author} {\bibfnamefont {J.}~\bibnamefont {Mai}}, \bibinfo {author} {\bibfnamefont {L.}~\bibnamefont {Zhang}}, \bibinfo {author} {\bibfnamefont {P.}~\bibnamefont {Zheng}}, \bibinfo {author} {\bibfnamefont {H.}~\bibnamefont {Yu}}, \bibinfo {author} {\bibfnamefont {C.-L.}\ \bibnamefont {Zou}}, \bibinfo {author} {\bibfnamefont {S.}~\bibnamefont {Liu}}, \bibinfo {author} {\bibfnamefont {F.}~\bibnamefont {Yan}}, \bibinfo {author} {\bibfnamefont {Y.}~\bibnamefont {Xu}}, \ and\ \bibinfo {author} {\bibfnamefont {D.}~\bibnamefont {Yu}},\ }\href {\doibase 10.1038/s41567-024-02619-5} {\bibfield  {journal} {\bibinfo  {journal} {Nature Physics}\ }\textbf {\bibinfo {volume} {20}},\ \bibinfo {pages}
  {1874} (\bibinfo {year} {2024})}\BibitemShut {NoStop}%
\bibitem [{\citenamefont {Fadel}\ \emph {et~al.}(2024)\citenamefont {Fadel}, \citenamefont {Roux},\ and\ \citenamefont {Gessner}}]{fadel2024reviewCV}%
  \BibitemOpen
  \bibfield  {author} {\bibinfo {author} {\bibfnamefont {M.}~\bibnamefont {Fadel}}, \bibinfo {author} {\bibfnamefont {N.}~\bibnamefont {Roux}}, \ and\ \bibinfo {author} {\bibfnamefont {M.}~\bibnamefont {Gessner}},\ }\href {https://arxiv.org/abs/2411.04122} {\enquote {\bibinfo {title} {Quantum metrology with a continuous-variable system},}\ } (\bibinfo {year} {2024}),\ \Eprint {http://arxiv.org/abs/2411.04122} {arXiv:2411.04122 [quant-ph]} \BibitemShut {NoStop}%
\bibitem [{\citenamefont {Oh}\ \emph {et~al.}(2020)\citenamefont {Oh}, \citenamefont {Park}, \citenamefont {Filip}, \citenamefont {Jeong},\ and\ \citenamefont {Marek}}]{Oh_NJP_2020}%
  \BibitemOpen
  \bibfield  {author} {\bibinfo {author} {\bibfnamefont {C.}~\bibnamefont {Oh}}, \bibinfo {author} {\bibfnamefont {K.}~\bibnamefont {Park}}, \bibinfo {author} {\bibfnamefont {R.}~\bibnamefont {Filip}}, \bibinfo {author} {\bibfnamefont {H.}~\bibnamefont {Jeong}}, \ and\ \bibinfo {author} {\bibfnamefont {P.}~\bibnamefont {Marek}},\ }\href {\doibase 10.1088/1367-2630/abd0b8} {\bibfield  {journal} {\bibinfo  {journal} {New Journal of Physics}\ }\textbf {\bibinfo {volume} {22}},\ \bibinfo {pages} {123039} (\bibinfo {year} {2020})}\BibitemShut {NoStop}%
\bibitem [{\citenamefont {G\'orecki}\ \emph {et~al.}(2022)\citenamefont {G\'orecki}, \citenamefont {Riccardi},\ and\ \citenamefont {Maccone}}]{WojciechPRL22}%
  \BibitemOpen
  \bibfield  {author} {\bibinfo {author} {\bibfnamefont {W.}~\bibnamefont {G\'orecki}}, \bibinfo {author} {\bibfnamefont {A.}~\bibnamefont {Riccardi}}, \ and\ \bibinfo {author} {\bibfnamefont {L.}~\bibnamefont {Maccone}},\ }\href {\doibase 10.1103/PhysRevLett.129.240503} {\bibfield  {journal} {\bibinfo  {journal} {Phys. Rev. Lett.}\ }\textbf {\bibinfo {volume} {129}},\ \bibinfo {pages} {240503} (\bibinfo {year} {2022})}\BibitemShut {NoStop}%
\bibitem [{\citenamefont {Filip}\ and\ \citenamefont {Mišta}(2011)}]{filip_detecting_2011}%
  \BibitemOpen
  \bibfield  {author} {\bibinfo {author} {\bibfnamefont {R.}~\bibnamefont {Filip}}\ and\ \bibinfo {author} {\bibfnamefont {L.}~\bibnamefont {Mišta}},\ }\href {\doibase 10.1103/PhysRevLett.106.200401} {\bibfield  {journal} {\bibinfo  {journal} {Phys. Rev. Lett.}\ }\textbf {\bibinfo {volume} {106}},\ \bibinfo {pages} {200401} (\bibinfo {year} {2011})}\BibitemShut {NoStop}%
\bibitem [{\citenamefont {Lachman}\ \emph {et~al.}(2019)\citenamefont {Lachman}, \citenamefont {Straka}, \citenamefont {Hloušek}, \citenamefont {Ježek},\ and\ \citenamefont {Filip}}]{lachman_faithful_2019}%
  \BibitemOpen
  \bibfield  {author} {\bibinfo {author} {\bibfnamefont {L.}~\bibnamefont {Lachman}}, \bibinfo {author} {\bibfnamefont {I.}~\bibnamefont {Straka}}, \bibinfo {author} {\bibfnamefont {J.}~\bibnamefont {Hloušek}}, \bibinfo {author} {\bibfnamefont {M.}~\bibnamefont {Ježek}}, \ and\ \bibinfo {author} {\bibfnamefont {R.}~\bibnamefont {Filip}},\ }\href {\doibase 10.1103/PhysRevLett.123.043601} {\bibfield  {journal} {\bibinfo  {journal} {Phys. Rev. Lett.}\ }\textbf {\bibinfo {volume} {123}},\ \bibinfo {pages} {043601} (\bibinfo {year} {2019})}\BibitemShut {NoStop}%
\bibitem [{\citenamefont {Podhora}\ \emph {et~al.}(2022)\citenamefont {Podhora}, \citenamefont {Lachman}, \citenamefont {Pham}, \citenamefont {Lešundák}, \citenamefont {Číp}, \citenamefont {Slodička},\ and\ \citenamefont {Filip}}]{podhora_quantum_2022}%
  \BibitemOpen
  \bibfield  {author} {\bibinfo {author} {\bibfnamefont {L.}~\bibnamefont {Podhora}}, \bibinfo {author} {\bibfnamefont {L.}~\bibnamefont {Lachman}}, \bibinfo {author} {\bibfnamefont {T.}~\bibnamefont {Pham}}, \bibinfo {author} {\bibfnamefont {A.}~\bibnamefont {Lešundák}}, \bibinfo {author} {\bibfnamefont {O.}~\bibnamefont {Číp}}, \bibinfo {author} {\bibfnamefont {L.}~\bibnamefont {Slodička}}, \ and\ \bibinfo {author} {\bibfnamefont {R.}~\bibnamefont {Filip}},\ }\href {\doibase 10.1103/PhysRevLett.129.013602} {\bibfield  {journal} {\bibinfo  {journal} {Phys. Rev. Lett.}\ }\textbf {\bibinfo {volume} {129}},\ \bibinfo {pages} {013602} (\bibinfo {year} {2022})}\BibitemShut {NoStop}%
\bibitem [{\citenamefont {Khaneja}\ \emph {et~al.}(2005)\citenamefont {Khaneja}, \citenamefont {Reiss}, \citenamefont {Kehlet}, \citenamefont {Schulte-Herbr{\"u}ggen},\ and\ \citenamefont {Glaser}}]{khaneja2005optimal}%
  \BibitemOpen
  \bibfield  {author} {\bibinfo {author} {\bibfnamefont {N.}~\bibnamefont {Khaneja}}, \bibinfo {author} {\bibfnamefont {T.}~\bibnamefont {Reiss}}, \bibinfo {author} {\bibfnamefont {C.}~\bibnamefont {Kehlet}}, \bibinfo {author} {\bibfnamefont {T.}~\bibnamefont {Schulte-Herbr{\"u}ggen}}, \ and\ \bibinfo {author} {\bibfnamefont {S.~J.}\ \bibnamefont {Glaser}},\ }\href@noop {} {\bibfield  {journal} {\bibinfo  {journal} {Journal of magnetic resonance}\ }\textbf {\bibinfo {volume} {172}},\ \bibinfo {pages} {296} (\bibinfo {year} {2005})}\BibitemShut {NoStop}%
\bibitem [{\citenamefont {Eickbusch}\ \emph {et~al.}(2022)\citenamefont {Eickbusch}, \citenamefont {Sivak}, \citenamefont {Ding}, \citenamefont {Elder}, \citenamefont {Jha}, \citenamefont {Venkatraman}, \citenamefont {Royer}, \citenamefont {Girvin}, \citenamefont {Schoelkopf},\ and\ \citenamefont {Devoret}}]{eickbusch2022fast}%
  \BibitemOpen
  \bibfield  {author} {\bibinfo {author} {\bibfnamefont {A.}~\bibnamefont {Eickbusch}}, \bibinfo {author} {\bibfnamefont {V.}~\bibnamefont {Sivak}}, \bibinfo {author} {\bibfnamefont {A.~Z.}\ \bibnamefont {Ding}}, \bibinfo {author} {\bibfnamefont {S.~S.}\ \bibnamefont {Elder}}, \bibinfo {author} {\bibfnamefont {S.~R.}\ \bibnamefont {Jha}}, \bibinfo {author} {\bibfnamefont {J.}~\bibnamefont {Venkatraman}}, \bibinfo {author} {\bibfnamefont {B.}~\bibnamefont {Royer}}, \bibinfo {author} {\bibfnamefont {S.~M.}\ \bibnamefont {Girvin}}, \bibinfo {author} {\bibfnamefont {R.~J.}\ \bibnamefont {Schoelkopf}}, \ and\ \bibinfo {author} {\bibfnamefont {M.~H.}\ \bibnamefont {Devoret}},\ }\href {\doibase 10.1038/s41567-022-01776-9} {\bibfield  {journal} {\bibinfo  {journal} {Nature Physics}\ }\textbf {\bibinfo {volume} {18}},\ \bibinfo {pages} {1464} (\bibinfo {year} {2022})}\BibitemShut {NoStop}%
\bibitem [{\citenamefont {Caneva}\ \emph {et~al.}(2009)\citenamefont {Caneva}, \citenamefont {Murphy}, \citenamefont {Calarco}, \citenamefont {Fazio}, \citenamefont {Montangero}, \citenamefont {Giovannetti},\ and\ \citenamefont {Santoro}}]{CanevaPRL09}%
  \BibitemOpen
  \bibfield  {author} {\bibinfo {author} {\bibfnamefont {T.}~\bibnamefont {Caneva}}, \bibinfo {author} {\bibfnamefont {M.}~\bibnamefont {Murphy}}, \bibinfo {author} {\bibfnamefont {T.}~\bibnamefont {Calarco}}, \bibinfo {author} {\bibfnamefont {R.}~\bibnamefont {Fazio}}, \bibinfo {author} {\bibfnamefont {S.}~\bibnamefont {Montangero}}, \bibinfo {author} {\bibfnamefont {V.}~\bibnamefont {Giovannetti}}, \ and\ \bibinfo {author} {\bibfnamefont {G.~E.}\ \bibnamefont {Santoro}},\ }\href {\doibase 10.1103/PhysRevLett.103.240501} {\bibfield  {journal} {\bibinfo  {journal} {Phys. Rev. Lett.}\ }\textbf {\bibinfo {volume} {103}},\ \bibinfo {pages} {240501} (\bibinfo {year} {2009})}\BibitemShut {NoStop}%
\bibitem [{\citenamefont {Hofheinz}\ \emph {et~al.}(2008)\citenamefont {Hofheinz}, \citenamefont {Weig}, \citenamefont {Ansmann}, \citenamefont {Bialczak}, \citenamefont {Lucero}, \citenamefont {Neeley}, \citenamefont {O’connell}, \citenamefont {Wang}, \citenamefont {Martinis},\ and\ \citenamefont {Cleland}}]{Hofheinz2008}%
  \BibitemOpen
  \bibfield  {author} {\bibinfo {author} {\bibfnamefont {M.}~\bibnamefont {Hofheinz}}, \bibinfo {author} {\bibfnamefont {E.}~\bibnamefont {Weig}}, \bibinfo {author} {\bibfnamefont {M.}~\bibnamefont {Ansmann}}, \bibinfo {author} {\bibfnamefont {R.~C.}\ \bibnamefont {Bialczak}}, \bibinfo {author} {\bibfnamefont {E.}~\bibnamefont {Lucero}}, \bibinfo {author} {\bibfnamefont {M.}~\bibnamefont {Neeley}}, \bibinfo {author} {\bibfnamefont {A.}~\bibnamefont {O’connell}}, \bibinfo {author} {\bibfnamefont {H.}~\bibnamefont {Wang}}, \bibinfo {author} {\bibfnamefont {J.~M.}\ \bibnamefont {Martinis}}, \ and\ \bibinfo {author} {\bibfnamefont {A.}~\bibnamefont {Cleland}},\ }\href {\doibase 10.1038/nature07136} {\bibfield  {journal} {\bibinfo  {journal} {Nature}\ }\textbf {\bibinfo {volume} {454}},\ \bibinfo {pages} {310} (\bibinfo {year} {2008})}\BibitemShut {NoStop}%
\bibitem [{SM()}]{SM}%
  \BibitemOpen
  \href@noop {} {\bibinfo  {journal} {See supplementary materials}\ }\BibitemShut {NoStop}%
\bibitem [{\citenamefont {Chabaud}\ \emph {et~al.}(2021)\citenamefont {Chabaud}, \citenamefont {Roeland}, \citenamefont {Walschaers}, \citenamefont {Grosshans}, \citenamefont {Parigi}, \citenamefont {Markham},\ and\ \citenamefont {Treps}}]{chabaud_certification_2021}%
  \BibitemOpen
\bibfield  {journal} {  }\bibfield  {author} {\bibinfo {author} {\bibfnamefont {U.}~\bibnamefont {Chabaud}}, \bibinfo {author} {\bibfnamefont {G.}~\bibnamefont {Roeland}}, \bibinfo {author} {\bibfnamefont {M.}~\bibnamefont {Walschaers}}, \bibinfo {author} {\bibfnamefont {F.}~\bibnamefont {Grosshans}}, \bibinfo {author} {\bibfnamefont {V.}~\bibnamefont {Parigi}}, \bibinfo {author} {\bibfnamefont {D.}~\bibnamefont {Markham}}, \ and\ \bibinfo {author} {\bibfnamefont {N.}~\bibnamefont {Treps}},\ }\href {\doibase 10.1103/PRXQuantum.2.020333} {\bibfield  {journal} {\bibinfo  {journal} {{PRX} Quantum}\ }\textbf {\bibinfo {volume} {2}},\ \bibinfo {pages} {020333} (\bibinfo {year} {2021})}\BibitemShut {NoStop}%
\bibitem [{\citenamefont {Fiur\'{a}\v{s}ek}(2022)}]{Fiurasek_22}%
  \BibitemOpen
  \bibfield  {author} {\bibinfo {author} {\bibfnamefont {J.}~\bibnamefont {Fiur\'{a}\v{s}ek}},\ }\href {\doibase 10.1364/OE.466175} {\bibfield  {journal} {\bibinfo  {journal} {Opt. Express}\ }\textbf {\bibinfo {volume} {30}},\ \bibinfo {pages} {30630} (\bibinfo {year} {2022})}\BibitemShut {NoStop}%
\bibitem [{\citenamefont {Provazn\'{i}k}\ \emph {et~al.}(2020)\citenamefont {Provazn\'{i}k}, \citenamefont {Lachman}, \citenamefont {Filip},\ and\ \citenamefont {Marek}}]{Provaznik20}%
  \BibitemOpen
  \bibfield  {author} {\bibinfo {author} {\bibfnamefont {J.}~\bibnamefont {Provazn\'{i}k}}, \bibinfo {author} {\bibfnamefont {L.}~\bibnamefont {Lachman}}, \bibinfo {author} {\bibfnamefont {R.}~\bibnamefont {Filip}}, \ and\ \bibinfo {author} {\bibfnamefont {P.}~\bibnamefont {Marek}},\ }\href {\doibase 10.1364/OE.389619} {\bibfield  {journal} {\bibinfo  {journal} {Opt. Express}\ }\textbf {\bibinfo {volume} {28}},\ \bibinfo {pages} {14839} (\bibinfo {year} {2020})}\BibitemShut {NoStop}%
\bibitem [{\citenamefont {Lee}(1991)}]{Lee_depth_PRA91}%
  \BibitemOpen
  \bibfield  {author} {\bibinfo {author} {\bibfnamefont {C.~T.}\ \bibnamefont {Lee}},\ }\href {\doibase 10.1103/PhysRevA.44.R2775} {\bibfield  {journal} {\bibinfo  {journal} {Phys. Rev. A}\ }\textbf {\bibinfo {volume} {44}},\ \bibinfo {pages} {R2775} (\bibinfo {year} {1991})}\BibitemShut {NoStop}%
\bibitem [{\citenamefont {Lee}(1992)}]{Lee_depth_PRA92}%
  \BibitemOpen
  \bibfield  {author} {\bibinfo {author} {\bibfnamefont {C.~T.}\ \bibnamefont {Lee}},\ }\href {\doibase 10.1103/PhysRevA.45.6586} {\bibfield  {journal} {\bibinfo  {journal} {Phys. Rev. A}\ }\textbf {\bibinfo {volume} {45}},\ \bibinfo {pages} {6586} (\bibinfo {year} {1992})}\BibitemShut {NoStop}%
\bibitem [{\citenamefont {Straka}\ \emph {et~al.}(2014)\citenamefont {Straka}, \citenamefont {Predojevi\ifmmode~\acute{c}\else \'{c}\fi{}}, \citenamefont {Huber}, \citenamefont {Lachman}, \citenamefont {Butschek}, \citenamefont {Mikov\'a}, \citenamefont {Mi\ifmmode~\check{c}\else \v{c}\fi{}uda}, \citenamefont {Solomon}, \citenamefont {Weihs}, \citenamefont {Je\ifmmode~\check{z}\else \v{z}\fi{}ek},\ and\ \citenamefont {Filip}}]{Radim_depth_14}%
  \BibitemOpen
  \bibfield  {author} {\bibinfo {author} {\bibfnamefont {I.}~\bibnamefont {Straka}}, \bibinfo {author} {\bibfnamefont {A.}~\bibnamefont {Predojevi\ifmmode~\acute{c}\else \'{c}\fi{}}}, \bibinfo {author} {\bibfnamefont {T.}~\bibnamefont {Huber}}, \bibinfo {author} {\bibfnamefont {L.}~\bibnamefont {Lachman}}, \bibinfo {author} {\bibfnamefont {L.}~\bibnamefont {Butschek}}, \bibinfo {author} {\bibfnamefont {M.}~\bibnamefont {Mikov\'a}}, \bibinfo {author} {\bibfnamefont {M.}~\bibnamefont {Mi\ifmmode~\check{c}\else \v{c}\fi{}uda}}, \bibinfo {author} {\bibfnamefont {G.~S.}\ \bibnamefont {Solomon}}, \bibinfo {author} {\bibfnamefont {G.}~\bibnamefont {Weihs}}, \bibinfo {author} {\bibfnamefont {M.}~\bibnamefont {Je\ifmmode~\check{z}\else \v{z}\fi{}ek}}, \ and\ \bibinfo {author} {\bibfnamefont {R.}~\bibnamefont {Filip}},\ }\href {\doibase 10.1103/PhysRevLett.113.223603} {\bibfield  {journal} {\bibinfo  {journal} {Phys. Rev. Lett.}\ }\textbf {\bibinfo {volume} {113}},\ \bibinfo {pages} {223603} (\bibinfo {year}
  {2014})}\BibitemShut {NoStop}%
\bibitem [{\citenamefont {Kwon}\ \emph {et~al.}(2019)\citenamefont {Kwon}, \citenamefont {Tan}, \citenamefont {Volkoff},\ and\ \citenamefont {Jeong}}]{KwonPRL19}%
  \BibitemOpen
  \bibfield  {author} {\bibinfo {author} {\bibfnamefont {H.}~\bibnamefont {Kwon}}, \bibinfo {author} {\bibfnamefont {K.~C.}\ \bibnamefont {Tan}}, \bibinfo {author} {\bibfnamefont {T.}~\bibnamefont {Volkoff}}, \ and\ \bibinfo {author} {\bibfnamefont {H.}~\bibnamefont {Jeong}},\ }\href {\doibase 10.1103/PhysRevLett.122.040503} {\bibfield  {journal} {\bibinfo  {journal} {Phys. Rev. Lett.}\ }\textbf {\bibinfo {volume} {122}},\ \bibinfo {pages} {040503} (\bibinfo {year} {2019})}\BibitemShut {NoStop}%
\bibitem [{\citenamefont {Sturges}\ \emph {et~al.}(2021)\citenamefont {Sturges}, \citenamefont {McDermott}, \citenamefont {Buraczewski}, \citenamefont {Clements}, \citenamefont {Renema}, \citenamefont {Nam}, \citenamefont {Gerrits}, \citenamefont {Lita}, \citenamefont {Kolthammer}, \citenamefont {Eckstein}, \citenamefont {Walmsley},\ and\ \citenamefont {Stobińska}}]{sturges_quantum_2021}%
  \BibitemOpen
  \bibfield  {author} {\bibinfo {author} {\bibfnamefont {T.~J.}\ \bibnamefont {Sturges}}, \bibinfo {author} {\bibfnamefont {T.}~\bibnamefont {McDermott}}, \bibinfo {author} {\bibfnamefont {A.}~\bibnamefont {Buraczewski}}, \bibinfo {author} {\bibfnamefont {W.~R.}\ \bibnamefont {Clements}}, \bibinfo {author} {\bibfnamefont {J.~J.}\ \bibnamefont {Renema}}, \bibinfo {author} {\bibfnamefont {S.~W.}\ \bibnamefont {Nam}}, \bibinfo {author} {\bibfnamefont {T.}~\bibnamefont {Gerrits}}, \bibinfo {author} {\bibfnamefont {A.}~\bibnamefont {Lita}}, \bibinfo {author} {\bibfnamefont {W.~S.}\ \bibnamefont {Kolthammer}}, \bibinfo {author} {\bibfnamefont {A.}~\bibnamefont {Eckstein}}, \bibinfo {author} {\bibfnamefont {I.~A.}\ \bibnamefont {Walmsley}}, \ and\ \bibinfo {author} {\bibfnamefont {M.}~\bibnamefont {Stobińska}},\ }\href {\doibase 10.1038/s41534-021-00427-w} {\bibfield  {journal} {\bibinfo  {journal} {npj Quantum Information}\ }\textbf {\bibinfo {volume} {7}},\ \bibinfo {pages} {91} (\bibinfo {year}
  {2021})}\BibitemShut {NoStop}%
\bibitem [{\citenamefont {Ritboon}\ \emph {et~al.}(2022)\citenamefont {Ritboon}, \citenamefont {Slodička},\ and\ \citenamefont {Filip}}]{ritboon_sequential_2022}%
  \BibitemOpen
  \bibfield  {author} {\bibinfo {author} {\bibfnamefont {A.}~\bibnamefont {Ritboon}}, \bibinfo {author} {\bibfnamefont {L.}~\bibnamefont {Slodička}}, \ and\ \bibinfo {author} {\bibfnamefont {R.}~\bibnamefont {Filip}},\ }\href {\doibase 10.1088/2058-9565/ac3c52} {\bibfield  {journal} {\bibinfo  {journal} {Quantum Science and Technology}\ }\textbf {\bibinfo {volume} {7}},\ \bibinfo {pages} {015023} (\bibinfo {year} {2022})}\BibitemShut {NoStop}%
\end{thebibliography}%

\clearpage
\newpage

\setcounter{page}{1}

\begin{widetext}

\section*{Supplemental material for ``Genuine quantum non-Gaussianity and metrological sensitivity of Fock states prepared in a mechanical resonator''}

\section{I. Sensitivity for unitary evolutions}

The maximum sensitivity achievable by a state $\hat{\rho}$ for sensing a parameter $\theta$ imprinted by a generator $\hat{G}$ through the unitary evolution $\hat{\rho}_\theta = e^{-i \theta \hat{G}} \hat{\rho} e^{i \theta \hat{G}} $ is given by the Quantum Fisher Information (QFI). For a pure state $\rho$, this is given by
\begin{equation}\label{suppeq:QFI}
    F_Q[\hat{\rho},\hat{G}] = 4 \text{Var}[\hat{G}]_{\hat{\rho}} \;,
\end{equation}
where $\text{Var}[\hat{G}]_{\hat{\rho}} = \avg{\hat{G}^2}_{\hat{\rho}} - \avg{\hat{G}}^2_{\hat{\rho}}$ is the variance.

The quantum Cramér-Rao (CR) theorem states that the uncertainty $(\Delta \theta_\text{est})^2$ in estimating the parameter $\theta$ by performing $\nu$ measurements on a state $\hat{\rho}$ is bounded as
\begin{equation}\label{supp:QCRB}
    (\Delta \theta_\text{est})^2 \geq (\Delta \theta)^2_\text{QCRB} \equiv \dfrac{1}{\nu F_Q[\hat{\rho},\hat{G}]} \;,
\end{equation}
where $(\Delta \theta)^2_\text{QCRB}$ is the so-called quantum Cramér-Rao bound (QCRB).

By definition, the QFI is the classical Fisher information (FI) maximized over all possible measurements, namely $F_Q[\hat{\rho},\hat{G}]=\max_{\{\hat{E}_m\}} F[\hat{\rho},\hat{G},\{\hat{E}_m\}]$, where $\{\hat{E}_m\}$ is the set of positive operator-valued measures (POVM) defining the quantum measurement and 
\begin{equation}\label{suppeq:FI}
    F[\hat{\rho},\hat{G},\{\hat{E}_m\}] \equiv F[P(m|\theta)] = \sum_{m} \dfrac{1}{P(m|\theta)} \left( \dfrac{\partial P(m|\theta)}{\partial \theta} \right)^2 
\end{equation}
is the FI associated with the probability distribution $P(m|\theta)=\text{Tr}[\hat{\rho}(\theta)\hat{E}_m]$. Therefore, for a specific choice of measurement, it is not guaranteed that the QCRB is saturated, since this is possible only for an optimal measurement choice.

\subsection{I.A. Displacements with a phase reference}

For displacement sensing, $\hat{G}$ is proportional to the phase-space quadrature perpendicular to the direction along which the displacement occurs, such that $e^{-i d \hat{G}}$ is equivalent to a displacement operator. Namely, if we define $\hat{G}(\phi)=\sqrt{2}(\hat{x}\sin\phi + \hat{p}\cos\phi)$, we have
\begin{equation}\label{supp:disp}
    e^{-i d \hat{G}} = e^{\alpha \aad - \alpha^\ast \aa} \equiv \hat{D}(\alpha) \;,
\end{equation}
where $\alpha=d e^{-i \phi}$.
From Eq.~\eqref{suppeq:QFI} we see that a Fock state $\ket{n}$ achieves a QFI for sensing the displacement amplitude $\theta$ of
\begin{equation}\label{suppeq:QFItheta}
    F_Q[\ket{n},\hat{G}(\phi)] = 4(2n+1) \;.
\end{equation}

We now show that the QCRB can be saturated by measurements of the Fock states' distribution. Let us consider the state $\rho_{\alpha}$ which results from displacing by $\alpha$ the initial state $\rho$, as $\rho_{\alpha} = \hat{D}(\alpha) \rho \hat{D}^\dagger(\alpha)$. We define the probability distribution $P_m(\alpha)\equiv P(m|\alpha)=\bra{m}\rho_{\alpha}\ket{m}$, which in the case of a $\rho$ that is diagonal in the Fock basis takes the form
\begin{align}
    P_m(\alpha) &= \bra{m}\rho_{\alpha}\ket{m} \notag\\
    &= \sum_{n} \bra{n}\rho\ket{n} |\bra{m}\hat{D}(\alpha)\ket{n}|^2 \notag\\
    &= \sum_{n} P_n \delta_{m,n}(\alpha) \;,
\end{align}
where $P_n=\bra{n}\rho\ket{n}$ and
\begin{equation}\label{suppeq:delta}
    \delta_{m,n}(\alpha) \equiv |\bra{m}\hat{D}(\alpha)\ket{n}|^2 = \dfrac{m!}{n!} e^{-|\alpha|^2} |\alpha|^{2(n-m)} \left( L_m^{n-m}[|\alpha|^2 ] \right)^2 \;,
\end{equation}
where $L_a^b[x]$ is the generalized Laguerre polynomial.
From the definition of the FI Eq.~\eqref{suppeq:FI}, we have
\begin{align}
    F[P_m(\alpha)] &= \sum_{m=0}^\infty \dfrac{1}{P_m(\alpha)} \left( \dfrac{d}{d \alpha} P_m(\alpha) \right)^2 \;, \notag\\
    & = \sum_{m=0}^\infty \dfrac{1}{\sum_{n} P_n \delta_{m,n}(\alpha)} \left( \sum_{n} P_n \dfrac{d}{d \alpha} \delta_{m,n}(\alpha) \right)^2 \label{suppeq:FIalpha}\;.
\end{align}
Using Eq.~\eqref{suppeq:delta} we obtain $F[P_m(\alpha)]=4(2n+1)$, which coincides with Eq.~\eqref{suppeq:QFItheta} meaning that measuring the Fock state distribution $P_m(\alpha)$ is sufficient to saturate the QCRB. Here, note that this is the FI for estimating the amplitude of $\alpha$, not the change in expectation value of a quadrature $\theta=\alpha \sqrt{2}$ as considered in other works \cite{fadel2024reviewCV}. Since the FI transforms under change of variables as
\begin{equation}
    F[P(y)] = F[P(x)] \left( \dfrac{d x}{d y}\right)^2 \;,
\end{equation}
the latter would in fact be $F[P_m(\theta)]=F[P_m(\alpha)]/2=2(2n+1)$. For the same reason, the FI for estimating $\overline{n}=|\alpha|^2$ is $F[P_m(|\alpha|^2)]=(1+2n)/2 |\alpha|^2$.

\subsection{I.B. Displacements without a phase reference}

When the displacement direction $\phi$ is random, or unknown, we can discuss two cases.

First, if the direction changes on a timescale that is slow compared to the measurement repetition rate, then one can introduce the average QFI as \cite{fadel2024reviewCV}
\begin{equation}
    F_Q^{\text{avg}}[\hat{\rho}] = \dfrac{1}{2\pi} \int_0^{2\pi} \dd\phi\, F_Q^{\text{avg}}[\hat{\rho},\hat{G}(\phi)] \;.
\end{equation}
With a constraint on the particle number in $\hat{\rho}$, namely a fixed $\overline{n}=\text{Tr}[\hat{\rho} \hat{a}^\dagger \hat{a}]$, it is possible to find that $F_Q^{\text{avg}}[\hat{\rho}] \leq 4(1+2\overline{n})$ \cite{fadel2024reviewCV}. Comparing this result with Eq.~\eqref{suppeq:QFItheta}, it is possible to conclude that this inequality can be saturated by Fock states. This result is perhaps intuitive, since Fock states are invariant under phase space rotations, as evident from their rotationally symmetric Wigner function. 

The second interesting case is when the displacement direction changes on a timescale comparable to or faster than the measurement repetition rate. In this scenario it is possible to introduce the phase-averaged displaced state \cite{podhora_quantum_2022,ritboon_sequential_2022}
\begin{equation}
    \hat{\rho}_{|\alpha|} = \dfrac{1}{2\pi} \int_0^{2\pi} \dd\phi\, \hat{D}(\alpha) \rho \hat{D}^\dagger(\alpha) \;.
\end{equation}
Note that this operation erases any matrix element of $\hat{\rho}$ that is off-diagonal in the Fock basis. It is thus possible to conclude that, for a fixed positive integer $\overline{n}$, Fock states achieve the maximum allowed sensitivity \cite{podhora_quantum_2022}.

\section{II. Density matrix evolution under energy relaxation ad dephasing}

Let us consider a density matrix expressed in the Fock basis as $\op{\rho}=\sum_{m,n} \rho_{m,n}(0) \ket{m}\bra{n}$. Under the action of a channel with loss rate $\kappa$ and dephasing rate $\gamma_\phi$, the master equation with collapse operators $\sqrt{\kappa}\aa$ and $\sqrt{2\gamma_\phi}\aad \aa$ can be solved analytically. For $\rho_{m,n}(t) = \bra{m} \op{\rho}(t) \ket{n}$, it gives
\begin{equation}\label{suppeq:anMEsol}
    \rho_{m,n}(t) = e^{-i\omega(m-n)t} e^{- (m+n) \frac{\kappa}{2} t } e^{- (m-n)^2 \gamma_\phi t } \sum_{\ell=0}^\infty \sqrt{{{n+\ell}\choose{\ell}}{{m+\ell}\choose{\ell}}}  (1-e^{-\kappa t})^\ell \rho_{m+\ell,n+\ell}(0) \;.
\end{equation}
In the experiment considered in this work $\kappa \gg \gamma_\phi$. Moreover, we will use Eq.~\eqref{suppeq:anMEsol} to compute the evolution of Fock states $\hat{\rho}=\ket{n}$ that have no off-diagonal contributions, therefore dephasing will not need to be considered.

\section{III. Investigations of Fock states sensitivity under relaxation}

The result Eq.~\eqref{suppeq:anMEsol} allows us to investigate the expected sensitivity of Fock states affected by energy relaxation.
We show in Fig.~\ref{suppfig:FvsAmp} the Fisher information Eq.~\eqref{suppeq:FIalpha} vs. displacement amplitude $\alpha$ achieved by a specific Fock state $\ket{n}$ after a decay time $t$. For an easier comparison, we show in Fig.~\ref{suppfig:FvsTime} the same data, but organized by time rather than by $n$. The Fisher information in the limit $\alpha\rightarrow 0$ is shown in Fig.~\ref{suppfig:FvsTime0amp} and its maximum over $\alpha$ is shown in Fig.~\ref{suppfig:FvsTimeMax}. 

From Fig.~\ref{suppfig:FvsTime0amp}, we see that for $10^{-4}\lesssim t/T_1\lesssim 10$, which is the range of highest experimental relevance, the FI in the limit $\alpha\rightarrow 0$ is especially well approximated by the analytic function $4(n+1)e^{-n t/T_1}$, where $\eta \equiv e^{-t/T_1}$. This implies that a Fock state $\ket{m}$ outperforms a Fock state $\ket{n}$ only if $\eta > \left(\frac{1+n}{1+m}\right)^{1/(m-n)}$.

\begin{figure}
    \centering
    \includegraphics[width=0.8\linewidth]{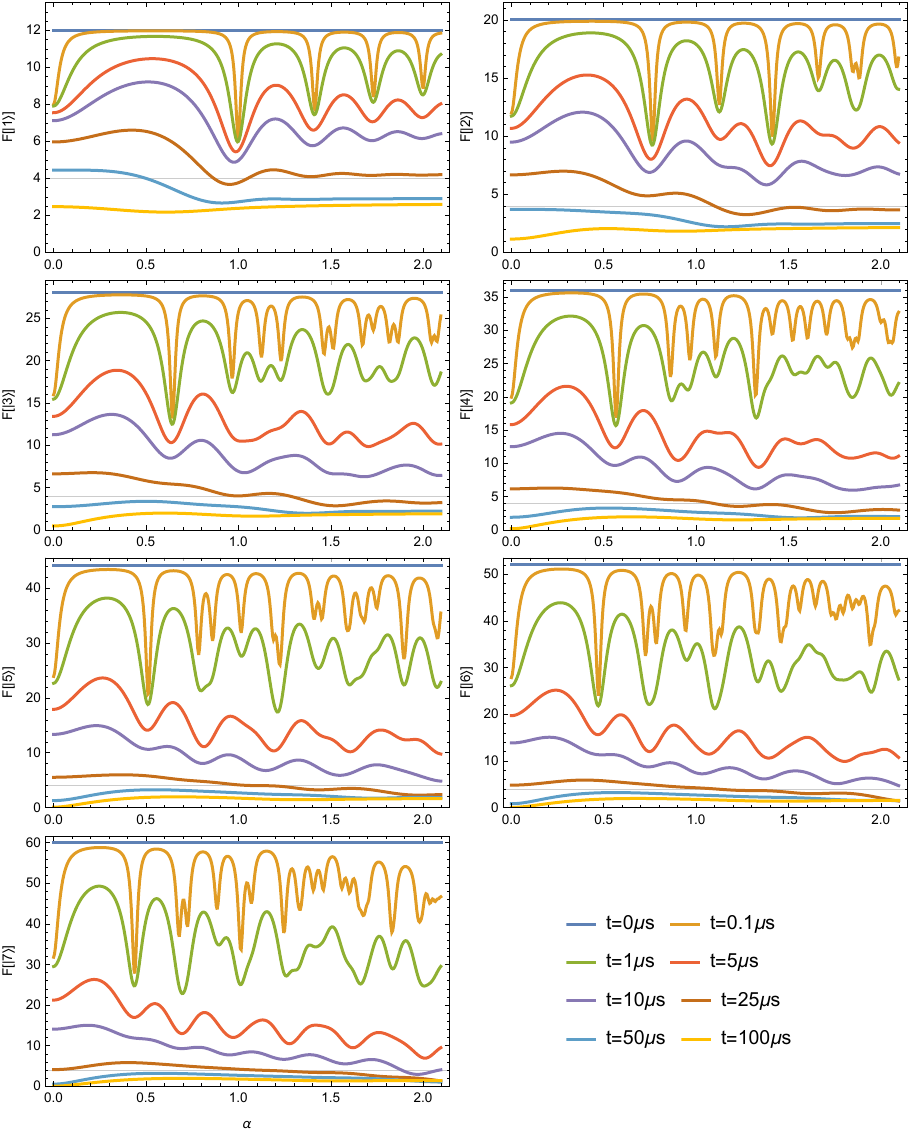}
    \caption{\textbf{Fisher information vs. displacement amplitude for different Fock states.} Each panel shows the Fisher information Eq.~\eqref{suppeq:FIalpha} vs. displacement amplitude $\alpha$ achieved by a specific Fock state $\ket{n}$ after a decay time $t$, see Eq.~\eqref{suppeq:anMEsol}. The decay rate is set to $\kappa=1/T_1=(\unit{85}{\mu s})^{-1}$. The horizontal blue line is the value $4(2n+1)$ for $t=0$, while the thin gray horizontal line at $4$ is the value achieved by a Fock $\ket{0}$ (or any other coherent state). See Fig.~\ref{suppfig:FvsTime} for a different visualization of the same data.}
    \label{suppfig:FvsAmp}
\end{figure}

\begin{figure}
    \centering
    \includegraphics[width=0.8\linewidth]{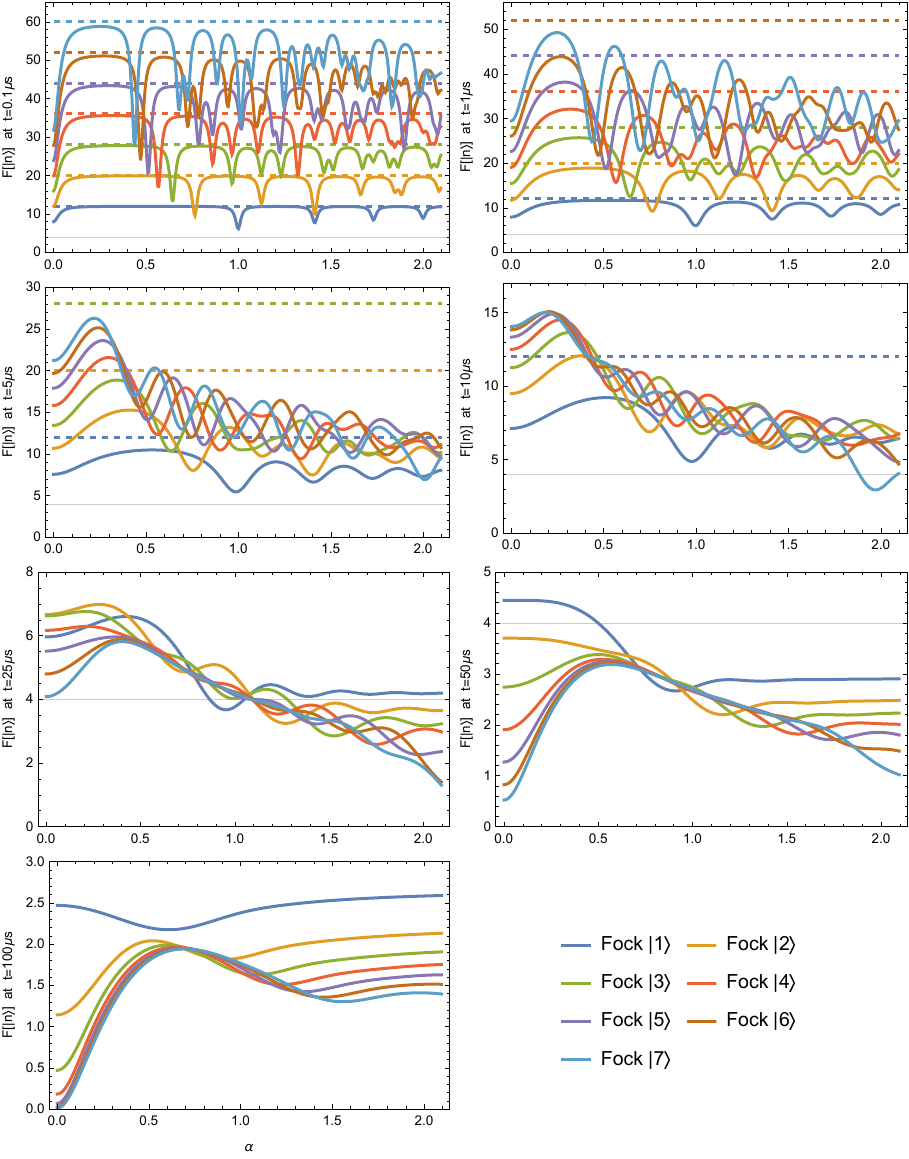}
    \caption{\textbf{Fisher information vs. displacement amplitude at different times.} Each panel shows the Fisher information Eq.~\eqref{suppeq:FIalpha} vs. displacement amplitude $\alpha$ achieved after a decay time $t$ by different Fock states $\ket{n}$. The colored curves are the same as the one shown in Fig.~\ref{suppfig:FvsAmp} but organized by time. The decay rate is set to $\kappa=1/T_1=(\unit{85}{\mu s})^{-1}$. The horizontal dashed lines are the value $4(2n+1)$ for $t=0$, while the thin gray horizontal line at $4$ is the value achieved by a Fock $\ket{0}$ (or any other coherent state).}
    \label{suppfig:FvsTime}
\end{figure}

\begin{figure}
    \centering
    \includegraphics[height=6.2cm]{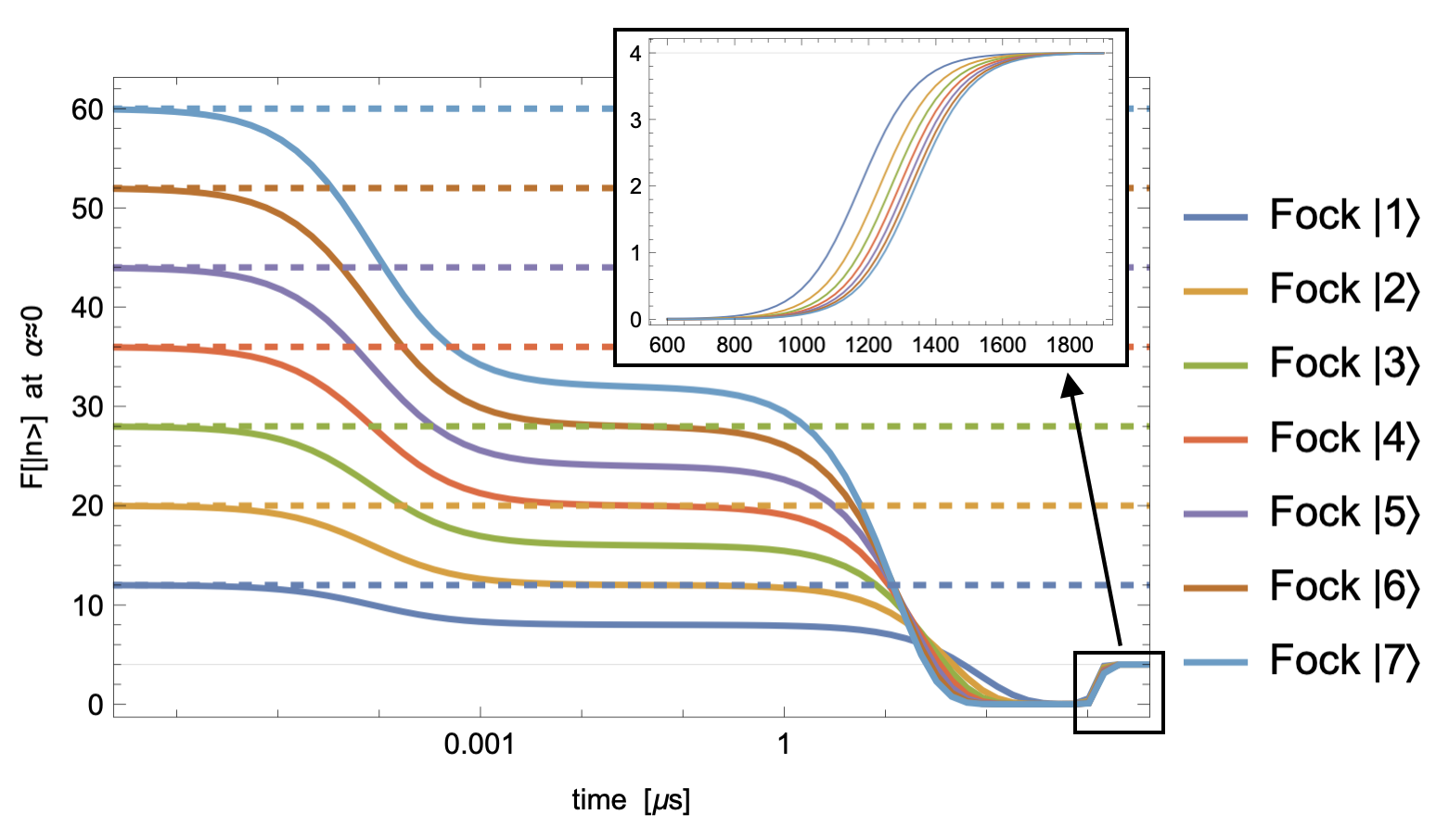}
    \includegraphics[height=5.8cm]{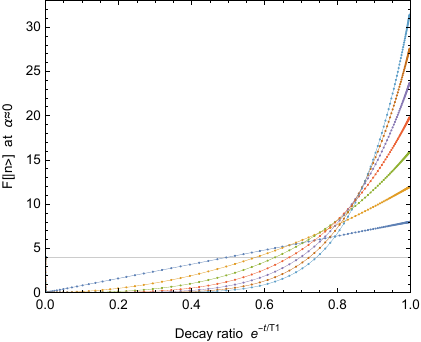}
    \caption{\textbf{Fisher information for small displacements at different times.} Left panel: In the limit $\alpha\rightarrow 0$, Fisher information Eq.~\eqref{suppeq:FIalpha} achieved after a decay time $t$ by different Fock states $\ket{n}$. The decay rate is set to $\kappa=1/T_1=(\unit{85}{\mu s})^{-1}$. The horizontal dashed lines are the value $4(2n+1)$ for $t=0$, while the thin gray horizontal line at $4$ is the value achieved by a Fock $\ket{0}$ (or any other coherent state).
    Note that the colored curves in this plot illustrate the behavior of the starting point (\ie $\alpha\approx 0$) of the colored curves shown in Figs.~\ref{suppfig:FvsAmp},\ref{suppfig:FvsTime}. It is interesting to note a rapid decay of the FI at $t/T_1 \sim 10^{-6}$, followed by a plateau around $10^{-5}\lesssim t/T_1 \lesssim 10^{-1}$. Then, at $t/T_1\approx 1$ the FI drops towards zero, to finally saturate at the value $4$ only for $t/T_1 > 10$.
    Right panel: Plotting the solid lines of the left panel as a function of $\eta \equiv e^{-t/T_1}$ shows that for $10^{-4}\lesssim t/T_1\lesssim 10$, which is the range of highest experimental relevance, the FI is especially well approximated by the analytic function $4(n+1)e^{-n t/T_1}$. Note here the absence of a factor 2 in front of $n$. This implies that a Fock state $\ket{m}$ outperforms a state $\ket{n}$ only if $\eta > \left(\frac{1+n}{1+m}\right)^{1/(m-n)}$, which is range above the point where the two corresponding lines intersects.
    }
    \label{suppfig:FvsTime0amp}
\end{figure}

\begin{figure*}
    \centering
    \includegraphics[height=5cm]{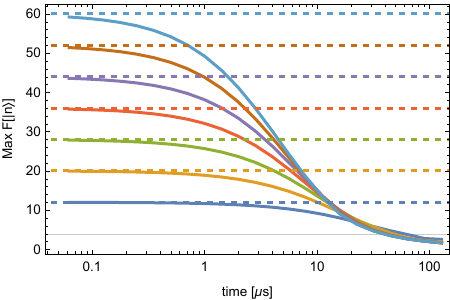}
    \includegraphics[height=5cm]{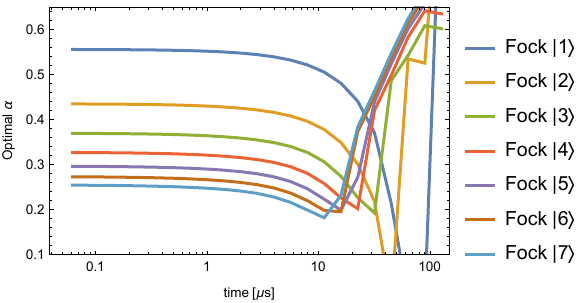}
    \caption{\textbf{Maximum Fisher information and optimal displacement amplitude at different times.} Left: maximum Fisher information Eq.~\eqref{suppeq:FIalpha} achieved after a decay time $t$ by different Fock states $\ket{n}$. The decay rate is set to $\kappa=1/T_1=(\unit{85}{\mu s})^{-1}$. The horizontal dashed lines are the value $4(2n+1)$ for $t=0$, while the thin gray horizontal line at $4$ is the value achieved by a Fock $\ket{0}$ (or any other coherent state).
    Right: value of the displacement amplitude $\alpha$ which maximizes the FI. 
    Note that the colored curves in these two plot illustrate the behavior of the maximum point of the colored curves shown in Figs.~\ref{suppfig:FvsAmp}, \ref{suppfig:FvsTime}. 
    }
    \label{suppfig:FvsTimeMax}
\end{figure*}

\clearpage
\newpage

\section{IV. Measured Fock State Distributions}

\begin{figure}[h!]
    \centering
    \includegraphics[width=0.77\linewidth]{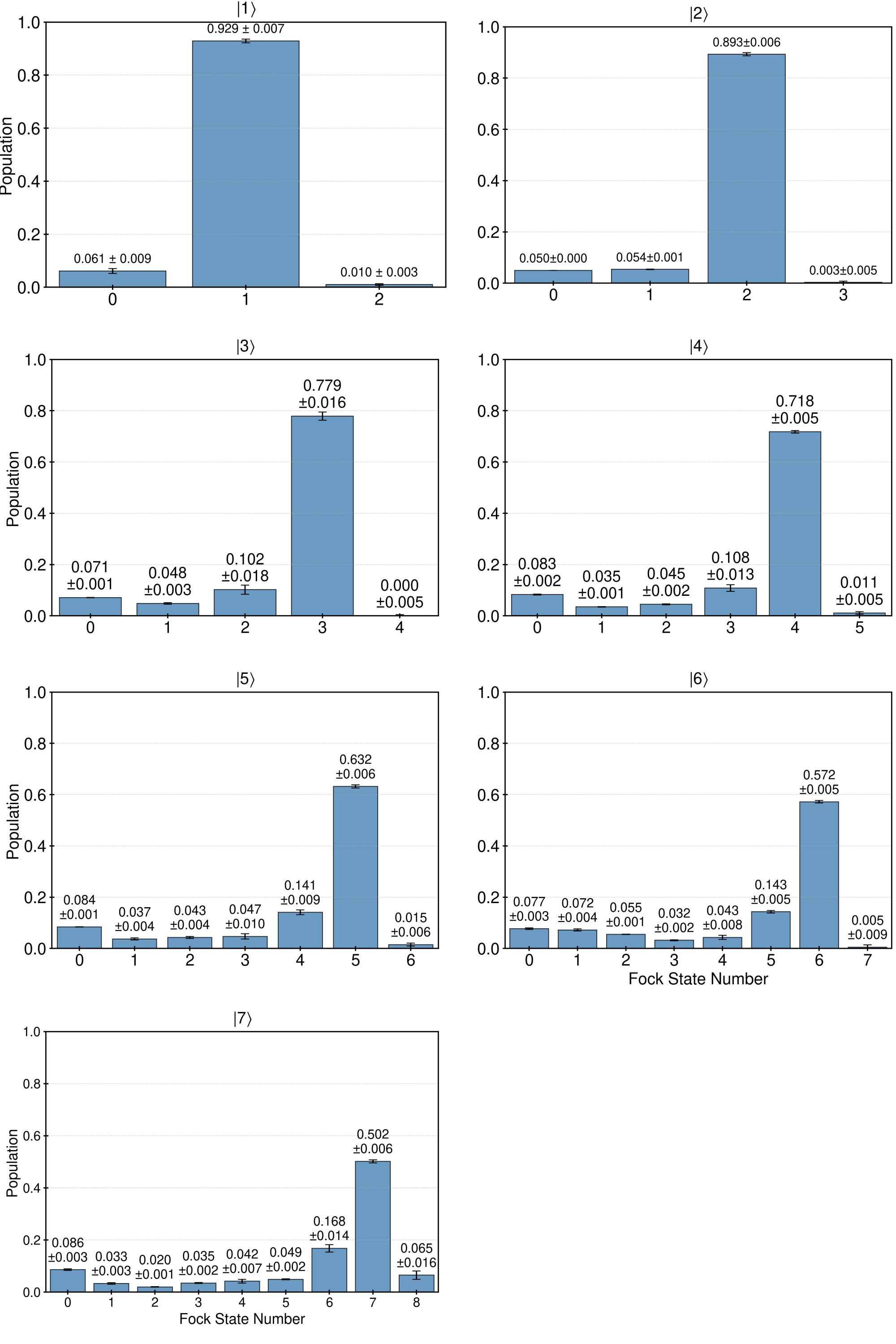}
    \caption{\textbf{Measured Fock state distributions.} For each of the experimentally prepared states, we show the Fock state number distributions as obtained from RPN measurement. Error bars come from fitting functions uncertainties. For $m>n+1$, the probabilities $P_m$ are zero within the measurement precision.}
    \label{suppfig:measFockDist}
\end{figure}

\section{V. Data analysis for displacement sensing}

\begin{figure}[h!]
    \centering
    \includegraphics[width=0.78\linewidth]{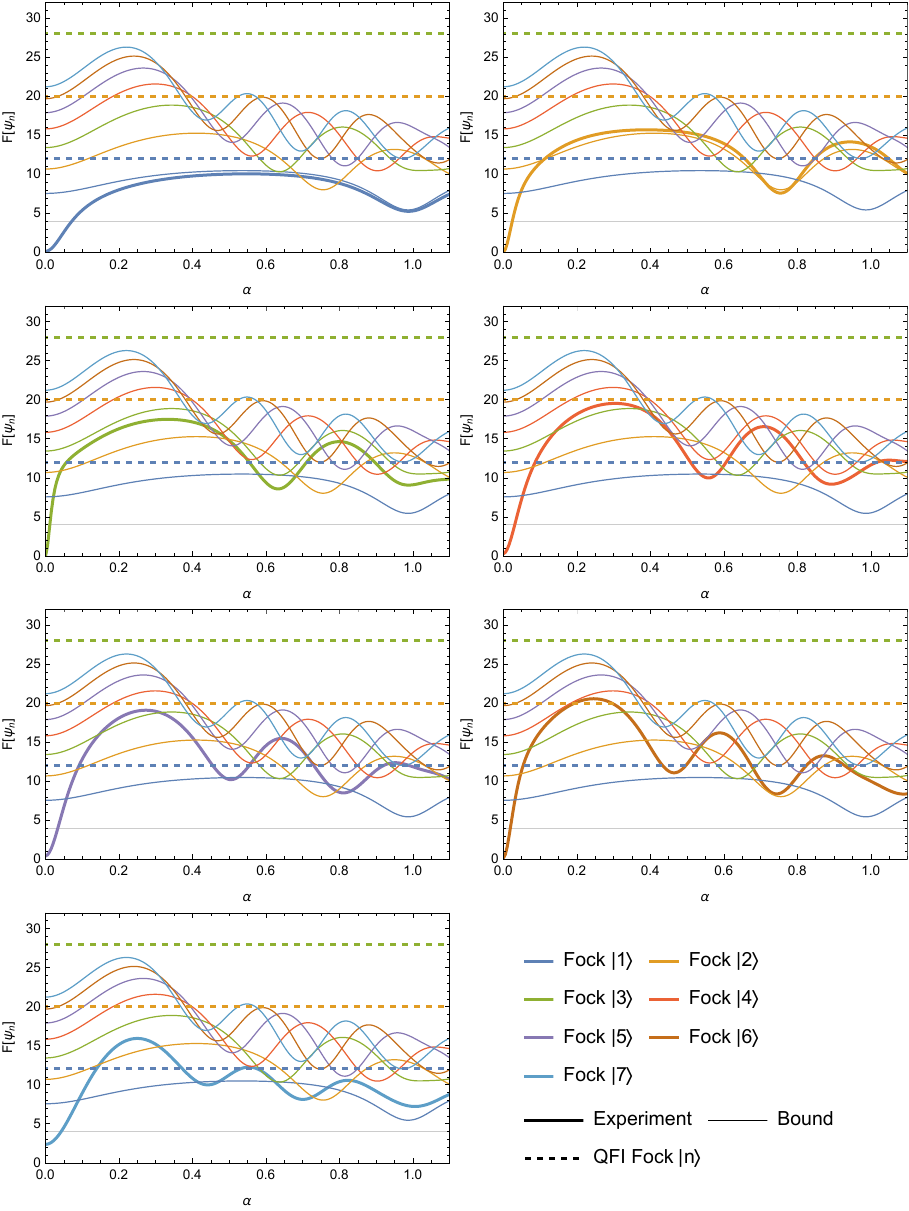}
    \caption{\textbf{Fisher information vs. displacement amplitude for the experimentally prepared states.} Each panel shows the Fisher information Eq.~\eqref{suppeq:FIalpha} vs. displacement amplitude $\alpha$ achieved after a decay time $t$ by different Fock states $\ket{n}$. 
    Thick curves are the FI for the experimentally prepared states, computed from the measured Fock states distributions shown in Fig.~\ref{suppfig:measFockDist}.
    Thin curves are the FI for ideal Fock states evolved for a time corresponding to our measurement time. The decay rate is set to $\kappa=1/T_1=(\unit{85}{\mu s})^{-1}$. The horizontal dashed lines are the value $4(2n+1)$ for $t=0$, while the thin gray horizontal line at $4$ is the value achieved by a Fock $\ket{0}$ (or any other coherent state).}
    \label{suppfig:FvsAmpData}
\end{figure}

\section{VI. Force sensitivity}

The Hamiltonian of a quantum harmonic oscillator with a time-dependent force applied reads
\begin{equation}\label{suppeq:driven_HO}
    \hat{H} = \dfrac{\hat{p}^2}{2 m} + \dfrac{1}{2} m \omega^2 \hat{x}^2 - F(t) \hat{x} \;.
\end{equation}
In terms of the bosonic creation and annihilation operators we have $\hat{x} = \sqrt{\frac{\hbar}{2m\omega}} \left(\aa + \aad \right)$, $\hat{p} = -i \sqrt{\frac{\hbar m\omega}{2}} \left(\aa - \aad \right)$, and the Hamiltonian Eq.~\eqref{suppeq:driven_HO} becomes
\begin{equation}
    \hat{H} = \underbrace{\hbar\omega \left( \aad \aa + \dfrac{1}{2} \right)}_{\hat{H}_0}  \underbrace{ - F(t) \xzpf \left(\aa + \aad \right) }_{\hat{H}_F} \;,
\end{equation}
where $\xzpf \equiv \sqrt{\hbar/2m\omega}$ is the zero point fluctuation of the oscillator. We now consider a resonant force $F(t)=F_0 \cos(\omega t + \phi) = \frac{F_0}{2} \left( e^{i(\omega t +\phi)} + e^{-i(\omega t +\phi)} \right)$ and move into the interaction picture with $\hat{U}=e^{-i \hat{H}_0 t /\hbar}$. Since $\hat{U}^\dagger \aa \hat{U} = \aa e^{-i\omega t}$, we obtain
\begin{align}
    \hat{H}_F' &= \hat{U}^\dagger \hat{H}_F \hat{U} \notag\\
    &=  - \dfrac{F_0 \xzpf}{2} \left( \aa e^{i\phi} + \aad e^{-i \phi} + \aa e^{-i(2\omega t+\phi)} + \aad e^{i(2\omega t+\phi)} \right) \\
    &\approx - \dfrac{F_0 \xzpf}{2} \left( \aa e^{i\phi} + \aad e^{-i \phi} \right) \;,
\end{align}
where in going to the second line we used the rotating wave approximation to neglect the fast rotating terms. Since under this approximation the Hamiltonian is time-independent, the time evolution of the system is given by
\begin{align}
    e^{-i (\hat{H}_F')^{\text{RWA}} t /\hbar} &= e^{ i \frac{F_0 \xzpf t}{2\hbar} \left( \aa e^{i\phi} + \aad e^{-i \phi} \right)}\\
    &= e^{\alpha \aad - \alpha^\ast \aa} \equiv \hat{D}(\alpha) \;,
\end{align}
namely the displacement operator with
\begin{equation}
    \alpha \equiv i \frac{F_0  \xzpf t}{2\hbar} e^{i\phi} \;.
\end{equation}

If we are interested in estimating $F_0$, \ie the modulo of the force, we can neglect any phase dependence and express the estimation uncertainty as
\begin{equation}
    \Delta F_0 = \dfrac{2 \hbar}{\xzpf t} \Delta \alpha \;.
\end{equation}
Given that the uncertainty $\Delta \alpha$ is bounded by the quantum Cramér-Rao bound Eq.~\eqref{supp:QCRB}, we obtain
\begin{equation}
    \Delta F_0 \geq \dfrac{2 \hbar}{\xzpf t} \dfrac{1}{\sqrt{\nu F_Q}} \;.
\end{equation}
Taking into account that to collect $\nu$ measurements requires a time $T = \nu t_\text{cycle} = \nu (t + t_\text{dead})$, with $t_\text{dead}$ the extra time needed to \eg initialize, readout and reset the probe system in each measurement shot, we can write
\begin{equation}
    \dfrac{\Delta F_0}{\sqrt{1/t_{\text{cycle}}}} \geq \dfrac{2 \hbar}{\xzpf} \dfrac{t_{\text{cycle}}}{t} \dfrac{1}{\sqrt{T F_Q}} \;.
\end{equation}
This last expression makes explicit the dependence of the sensitivity on the inverse of the duty cicle, $t_{\text{cycle}}/t$, meaning that the smaller $t_\text{dead}$ the better, and on the total measurement time $T$.

To make a concrete estimate, let us consider for the oscillator an effective mass of $m=\unit{16.2}{\mu g}$ \cite{catSCI23}, resulting in $\xzpf = \unit{3.2\cdot 10^{-19}}{m}$, a measurement time $t=\unit{90}{\mu s} \simeq T_1$, and a dead time $t_\text{dead} = \unit{210}{\mu s}$ that includes state preparation ($\sim\unit{2}{\mu s}$), RPN measurement time ($\sim\unit{10}{\mu s}$) and qubit/resonator reset ($\sim\unit{200}{\mu s}$). For $T=t_\text{cycle}$ and $F_Q=4$, which is the value for the standard quantum limit, we obtain a force sensitivity of $\unit{63.2}{fN/\sqrt{\text{Hz}}}$.

\end{widetext}

\end{document}